\documentclass[10pt,twocolumn,twoside]{IEEEtran} 
\usepackage{ifpdf}

\usepackage{cite}

\ifCLASSINFOpdf
   \usepackage[pdftex]{graphicx}
   \graphicspath{{./figs/pdf/}{./figs/jpeg/}}
   \DeclareGraphicsExtensions{.pdf,.jpeg,.png}
\else
   \usepackage[dvips]{graphicx}
   \graphicspath{{./figs/eps/}}
   \DeclareGraphicsExtensions{.eps}
\fi
\usepackage[cmex10]{amsmath,mathtools}
\usepackage{array}

\ifCLASSOPTIONcompsoc
  \usepackage[caption=false,font=normalsize,labelfont=sf,textfont=sf]{subfig}
\else
  \usepackage[caption=false,font=footnotesize]{subfig}
\fi
\usepackage{fixltx2e}
\usepackage{longtable}          
\usepackage{multirow}
\usepackage{framed}
\usepackage[table]{xcolor}
\usepackage{epstopdf}
\usepackage{amssymb}
\usepackage{enumerate}
\usepackage{dsfont}

\definecolor{CorCol}{rgb}{1,0,0}

\newcommand{\tran}{^{\mathsf{T}}}
\newcommand{\expectx}[1]{\mathds{E}_{\mathbf{x}_k}\!{\left\lbrace #1 \right\rbrace}}

\begin{document}
\allowdisplaybreaks
%
\title{Approximate MMSE Estimator for Linear Dynamic Systems with Gaussian Mixture Noise}
%
%
%

\author{Leila~Pishdad,
        Fabrice~Labeau
\thanks{The authors are with the Department of Electrical and Computer Engineering, McGill University, Montreal, QC H3A 0E9, Canada (e-mail: leila.pishdad@mail.mcgill.ca; fabrice.labeau@mcgill.ca)}
}

\maketitle

\begin{abstract}
In this work we propose an approximate Minimum Mean-Square Error (MMSE) filter for linear dynamic systems with Gaussian Mixture noise. The proposed estimator tracks each component of the Gaussian Mixture (GM) posterior with an individual filter and minimizes the trace of the covariance matrix of the bank of filters, as opposed to minimizing the MSE of individual filters in the commonly used Gaussian sum filter (GSF). Hence, the spread of means in the proposed method is smaller than that of GSF which makes it more robust to removing components. Consequently, lower complexity reduction schemes can be used with the proposed filter without losing estimation accuracy and precision. This is supported through simulations on synthetic data as well as experimental data related to an indoor localization system. Additionally, we show that in two limit cases the state estimation provided by our proposed method converges to that of GSF, and we provide simulation results supporting this in other cases.
\end{abstract}
%
%
%
%
%
%
%
%
%
%
\section{Introduction}
The problem of estimating the unobservable state of a dynamic system from its available noisy measurements is prevalent in numerous signal processing contexts.
Bayesian tracking techniques have been used for this purpose by using a probabilistic framework and approximating the posterior or \textit{belief function}, i.e. the conditional probability density function of the state given the measurements. For the special case of Gaussian noise with linear dynamic and measurement models, this posterior is Gaussian and its sufficient statistics are optimally tracked by a Kalman filter~\cite{ho_bayesian_1964,arulampalam_tutorial_2002}. The mean of this pdf acts as the state estimate and it is proved to be the minimum mean-square error (MMSE) estimator~\cite{anderson1979optimal}. However, for the case of non-Gaussian noise processes, approximations should be made to provide suboptimal solutions~\cite{arulampalam_tutorial_2002}.

Gaussian sum approximation has been an attractive method for estimating non-Gaussian distributions, since it provides asymptotically unbiased estimations~\cite{1962KernelDensityEstimation}, with the desired precision\footnote{The integral of the approximation error over the sample space can be made as small as desired.}~\cite[Chapter8; Lemma 4.1]{anderson1979optimal}. 
Additionally, by using Gaussian mixtures (GM), the approximated pdf is represented as a conditionally Gaussian distribution and this enables the analytic evaluation of a closed-form expression for the belief function. This is possible since, with GM distributions, the \textit{multiple model} approach can be used, where each component in the GM corresponds to a model in the system and can be tracked by a Kalman filter~\cite{bar2001estimation}. Hence, the partitioned posterior can be estimated by a bank of  Kalman filters, i.e. the Gaussian sum filter (GSF). Consequently, GMs have been widely used to model the different non-Gaussian distributions in sequential Bayesian tracking, including the prior ~\cite{anderson1979optimal,alspach_nonlinear_1972,sorenson_recursive_1971,flament_particle_2004,lehn-schioler_parzen_2004,bilik_optimal_2005}, likelihood~\cite{huber_hybrid_2007,huber_efficient_2007}, predictive density~\cite{huber_hybrid_2007_Other,stordal_bridging_2011,ali-loytty_efficient_2008}, and noise distributions~\cite{bar2001estimation,GSPFkotecha_gaussian_2003,chen_mixture_2000,alspach_nonlinear_1972,sorenson_recursive_1971,schoenberg_posterior_2012,bilik_mmse-based_2010,sun_mixture_2004,ali-loytty_box_2010,faubel_split_2009,bilik_optimal_2005}. They have also been used to directly approximate the posterior distribution~\cite{ito_gaussian_2000,GPFkotecha_gaussian_2003,crisan_generalised_2013,schoenberg_posterior_2012,terejanu_adaptive_2011,ali-loytty_efficient_2008,faubel_split_2009,schrempf_efficient_2007}.

With GM prior, likelihood, or predictive density, the posterior is also a GM and the number of its components remains constant over time, as long as the noise distributions are Gaussian. For instance, in~\cite{anderson1979optimal}, it is shown that starting with a GM prior\footnote{with a finite number of components}, and additive white Gaussian noise, the predictive and posterior distributions will also be GMs with the same number of components. However, for GM noise distributions, the number of models in the system and consequently the number of components in the posterior grow exponentially over time. Hence, suitable Gaussian Mixture reduction algorithms should be used, to merge or remove some of the components in the posterior as time progresses.

The mixture reduction algorithms can be categorized into three classes. In the first group, Expectation maximization (EM) is used to simultaneously predict and reduce the GM~\cite{huber_efficient_2007,bilik_optimal_2005,bilik_mmse-based_2010}, e.g. by running the EM algorithm on synthetically generated data~\cite{bilik_optimal_2005,bilik_mmse-based_2010}. The second class of reduction algorithms rely on merging a pair or a group of components, i.e. replacing them by their moment-matching Gaussian distribution. There are different criteria for selecting the components to be merged. For instance, in Gaussian pseudo Bayesian (GPB) estimators, the components with the same state history are merged and replaced by a single Gaussian distribution~\cite{bar2001estimation}. A less computationally complex solution, also approximating a GM by a single Gaussian, 
is interacting multiple models (IMM)~\cite{bar2001estimation} which is used commonly as it requires fewer filters. Alternatively, in~\cite{faubel_split_2009} the components are merged in the unlikely regions of the distribution, and they are split in the likely regions. In~\cite{west_approximating_1993}, the component with the smallest weight is merged with its closest neighbor.\footnote{The neighboring components have the smallest Euclidean distance between their means.}  Optimization techniques can also be used to select the merging components such that a cost function which quantifies the dissimilarity between the GM distribution and the reduced distribution is minimized. Some of the cost functions used in the literature include Kullback-Leibler divergence (KLD)~\cite{schoenberg_posterior_2012,runnalls_kullback-leibler_2007},  and integral square difference\footnote{This criterion is also referred to as integral square error (ISE)} (ISD)~\cite{schieferdecker_gaussian_2009,williams_cost-function-based_2003}. Alternatively in~\cite{salmond_mixture_1990} the authors merge the components leading to the minimum increase in the \textit{within-component} variance of the reduced GM distribution.
The last category of reduction schemes requires removing a group of components~\cite{alspach_nonlinear_1972,sorenson_recursive_1971}. This class has the lowest computational complexity, especially since the reduction can be done before tracking, making it possible to avoid the evaluation of unused parameters. In the extreme case, the active component of the posterior is determined and all the other components are removed~\cite{chen_mixture_2000,sun_mixture_2004}. This is equivalent to making a hard decision about the model in effect and hence the accuracy and precision of estimation is dependent on the correct choice of this model. At the cost of increased computational complexity, the performance of these methods can be improved by applying \textit{resampling} procedures~\cite{GSPFkotecha_gaussian_2003,crisan_generalised_2013,andrieu_particle_2002}. Alternatively, in~\cite{ali-loytty_box_2010,ali-loytty_efficient_2008} a \textit{forgetting and merging} algorithm is proposed, where the components with weights smaller than a given threshold are removed, and the components with close enough moments are merged. A comprehensive review of the reduction algorithms for GM distributions is provided in~\cite{crouse_look_2011}.


The MMSE estimator is the expected value of the posterior~\cite{poor_introduction_1994,bar2001estimation}. Hence, GSF is the MMSE estimator of state\footnote{If no reduction scheme is used and the distributions are GMs (not \textit{approximated} by GMs).}~\cite{ackerson_state_1970,tugnait_adaptive_1980,bilik_mmse-based_2010}. However, due to using parallel Kalman filters in GSF, the MSE of each individual filter is minimized irrespective of the location of its mean with respect to the other components. Yet, the total covariance matrix of the filter is a function of both the individual filters' state estimation covariance matrices and the spread of their means (as we show in~\eqref{eq:mergedCov}--\eqref{eq:spreadOfMeans} in Section~\ref{sec:GSF}).

In this work, we propose an approximate MMSE state estimator called AMMSE, which, unlike GSF, minimizes the trace of the total covariance matrix of the filter rather than the traces of the covariance matrices of individual filters. For this purpose, we re-derive the gains of individual filters such that the trace of the total covariance matrix of the filter is minimized. Hence, the spread of the means of the posterior components in the proposed estimator is smaller than that of GSF, making it more robust to removing some of the components. Consequently, AMMSE estimator can be used with the simplest and least computationally complex reduction scheme and yet achieve better performance when compared with GSF with the same reduction scheme. In other words, rather than using more computationally complex reduction schemes to improve the performance, we adjust the means of the components of the posterior and simply use the component with the largest weight as the state estimator, thus avoiding the unnecessary evaluation of the parameters of the other components. Additionally, through simulations we show  that the difference between the distributions of the estimated state in GSF and AMMSE, as well as the true state is not statistically significant. Hence, despite the fact that by changing the gains of individual filters we are deviating from the true posterior, the total state estimation of AMMSE converges to the MMSE state estimation provided by GSF, and the true state.

%
%

The rest of this paper is organized as follows: In Section~\ref{sec:Notations} the system model is defined and the notations used throughout the paper are introduced. Next, in Section~\ref{sec:GSF}, we provide the details of GSF. In Section~\ref{sec:myCont}, we present our proposed approximate MMSE filter: first, in Section~\ref{sec:DerivingGainsOptimal} the gains of individual filters in the proposed method are derived, and then in Section~\ref{sec:CompareGSFMMSE} it is compared  with GSF in terms of computational complexity and the convergence of the two filters is analyzed. The simulation results are provided in Section~\ref{sec:SimulationResults}, comparing AMMSE, GSF, Kalman and Matched filters in terms of estimation accuracy and precision with synthetic data (Section~\ref{sec:SimSynthetic}) and experimental data gathered from an indoor localization system (Section~\ref{sec:simExperimental}), and the convergence\footnote{Convergence in distribution is tested using two-sample Kolmogrov-Smirnov non-parametric tests.} of the state estimations of the filters and the true state is tested with synthetic data. Finally, in Section~\ref{sec:Conclusion} we provide concluding remarks. 
%

\section{System Model}
\label{sec:Notations}
Suppose a discrete-time linear dynamic system, in which the state sequence \(\{\mathbf{x}_{k},k \in  \mathbb{N}\}\), evolves as a first-order Markov process with additive noise. Hence, using state-space representation, the dynamics equation can be written as
\begin{align}
\label{eq:process}
\mathbf{x}_{k} & = F_{k}\mathbf{x}_{k-1}+\mathbf{v}_{k},
\end{align}
where the process noise, \(\{\mathbf{v}_{k},k \in  \mathbb{N}\}\) is an i.i.d. random vector sequence with the pdf \(p{ \left( \mathbf{v}_{k} \right)}\) and \(F_{k}\) is a known matrix which describes the linear relationship between the previous and current state. If \(n_x\) denotes the dimension of the state vector, the process noise is of dimension \(n_x\) and matrix \(F_k\) is of size \(n_x \times n_x\).

In many applications, the state of the system cannot be observed directly. Hence, it is desirable to estimate the unobservable state from the available measurements. If we denote the measurement sequence by \(\{\mathbf{z}_{k},k \in  \mathbb{N}\}\),  the relationship between \(\mathbf{x}_k\) and \(\mathbf{z}_k\) is described by the measurement equation,
\begin{align}
\label{eq:measurement}
\mathbf{z}_k & = H_k\mathbf{x}_{k} + \mathbf{w}_k,
\end{align}
where the measurement noise \(\{\mathbf{w}_{k},k \in  \mathbb{N}\}\) is an i.i.d. random vector sequence with the pdf  \(p {\left( \mathbf{w}_{k} \right)}\), and it is independent from the process noise. The matrix defining the linear relationship between the current state and measurement vectors, \(H_{k}\), is a known matrix of size \(n_z \times n_x\), where \(n_z\) is the dimension of the measurement vector \(\mathbf{z}_k\), and the noise vector \(\mathbf{w}_k\).

Having the above, Bayesian tracking techniques can be used to probabilistically estimate the current state of the system from the available measurements. This is done by recursively estimating the posterior, \(p {\left(\mathbf{x}_k|\mathbf{z}_{1:k}\right)} \), where \(\mathbf{z}_{1:k}\) represents the available measurements up to and including time \(k\). The recursive estimation of the posterior comprises of two steps: prediction and update. In the first step, \(\mathbf{x}_k\) is predicted using the previous measurements, i.e. the pdf \(p {\left(\mathbf{x}_{k}|\mathbf{z}_{1:k-1}\right)} \) is estimated using Chapman-Kolmogrov equation on the previous posterior, \(p {\left(\mathbf{x}_{k-1}|\mathbf{z}_{1:k-1}\right)}\), and \(p {\left(\mathbf{x}_{k}|\mathbf{x}_{k-1}\right)}\). Next, in the update phase, Bayes rule is used to update the prior and evaluate the posterior, \(p {\left(\mathbf{x}_{k}|\mathbf{z}_{1:k}\right)}\). The posterior is then used for next iteration estimation. With Gaussian process and measurement noise, the posterior distribution will be Gaussian and its sufficient statistics, i.e. mean and covariance matrix, are optimally tracked by a Kalman filter~\cite{ho_bayesian_1964}. Additionally, since the posterior is Gaussian, the filtered mean
approximates the state
and it is the MMSE estimator~\cite{poor_introduction_1994,bar2001estimation}. 

In this work we use GM models for process and measurement noise processes, since they are mathematically tractable and they can be used to approximate any non-Gaussian distribution. Specifically, with GM approximation, the pdf can be approximated as closely as desired~\cite[Lemma 4.1]{anderson1979optimal} and this estimation is asymptotically unbiased~\cite{1962KernelDensityEstimation}. Hence, the process noise, \(\mathbf{v}_k\) is estimated by a GM with \(C_{v_k}\) components, with \(\left\lbrace \mathbf{u}^i_k,1 \leq i \leq C_{v_k}\right\rbrace  \) the component means, \(\left\lbrace Q^i_k,1 \leq i \leq C_{v_k}\right\rbrace  \) the component covariance matrices and \(\left\lbrace w^i_k,1 \leq i \leq C_{v_k}\right\rbrace  \) the non-negative mixing coefficients. This can be written as:
\begin{align}
\label{eq:priorDist}
p {\left(\mathbf{v}_k\right)} \approx \sum \limits _{i=1}^{C_{v_k}} w^i_k \mathcal{N}{\left( \mathbf{v}_k; \mathbf{u}^i_k, Q^i_k \right)}, 
\end{align}
where \(\sum _{i=1}^{C_{v_k}} w^i_k=1\), and \(\mathcal{N}{\left( \mathbf{x};\boldsymbol{\mu},\Sigma \right)} \) represents a Gaussian distribution with argument \(\mathbf{x}\), mean \(\boldsymbol{\mu}\), and covariance matrix \(\Sigma\). The measurement noise distribution can also be approximated by a GM in a similar manner, and written as:
\begin{align}
\label{eq:likelihoodDist}
p {\left(\mathbf{w}_k\right)} \approx \sum \limits _{j=1}^{C_{w_k}} p^j_k \mathcal{N}{\left( \mathbf{w}_k; {\mathbf{b}}^j_k, R^j_k \right)},
\end{align}
where, \(C_{w_k}\) is the number of components of the GM distribution with the non-negative coefficients \(\left\lbrace p^j_k,1 \leq j \leq C_{w_k}\right\rbrace  \), and \(\sum  _{i=1}^{C_{w_k}} p^j_k=1\). The mean and covariance matrix of component \( j,1 \leq j \leq C_{w_k}\) are \({\mathbf{b}}^j_k\) and \(R^j_k\), respectively. 

\section{Gaussian Sum Filters}
\label{sec:GSF}
%
%
With the GM noise distributions, i.e.~\eqref{eq:priorDist}--\eqref{eq:likelihoodDist}, the dynamic system defined in~\eqref{eq:process}--\eqref{eq:measurement}, can be described as a \textit{Multiple Model} system, with models \(\left\lbrace M_{k}^{ij}; 1 \leq i \leq {C_{v_k}}, 1\leq j \leq {C_{w_k}} \right\rbrace\), corresponding to the different components or \textit{modes} of the process and measurement noises.\footnote{For simplicity we assume a system of order one, where all the models have the same history of states. This assumption can be easily relaxed.} Hence, the posterior can be partitioned
as follows:
\begin{align}
\label{eq:GSFmergedposterior}
p {\left(\mathbf{x}_k | \mathbf{z}_{1:k}\right)} & = \sum \limits _{i,j} p {\left(\mathbf{x}_k | \mathbf{z}_{1:k},M_{k}^{ij}\right)}p{ \left(M_{k}^{ij}|\mathbf{z}_{1:k}\right)}.
\end{align}

The mode-conditioned posterior, \(p {\left(\mathbf{x}_k |M_{k}^{ij}, \mathbf{z}_{1:k}\right)}\), is a Gaussian distribution with the pdf,
\begin{align}
\label{eq:individualKFposterior}
p {\left( \mathbf{x}_k|M_{k}^{ij},\mathbf{z}_{1:k} \right)} & = \mathcal{N} {\left(\mathbf{x}_k; \mathsf{\hat{\boldsymbol{x}}}_{k|k}^{ij}, \mathsf{P}_{k|k}^{ij}\right)}.
\end{align}
where its parameters \(\mathsf{\hat{\boldsymbol{x}}}_{k|k}^{ij}\) and \(\mathsf{P}_{k|k}^{ij}\) can be tracked using the \textit{mode-matched} Kalman filter~\cite{ho_bayesian_1964,bar2001estimation} as follows: 
\begin{align}
\label{eq:ijpredictedState}
\mathbf{\hat{x}}_{k|k-1}^{i} &= F_{k}\mathbf{\hat{x}}_{k-1|k-1} + \mathbf{u}_k^{i},
\\ \label{eq:ijpredictedCov} P_{k|k-1}^{i} & = Q_k^{i}+F_kP_{k-1|k-1}F_k\tran,
\\ \label{eq:ijpredictedInnov} \boldsymbol{\nu}_{k}^{ij} & = \mathbf{z}_{k} - \left(H_k\mathbf{\hat{x}}_{k|k-1}^{i} + \mathbf{b}_k^{j}\right),
\\ \label{eq:ijInnovationCov} S_k^{ij} & = H_k P_{k|k-1}^{i}H_k\tran + R_k^j,
\\ \label{eq:ijKalGain} \mathsf{W}_k^{ij} &= P_{k|k-1}^{i}H_k\tran{S_k^{ij}}^{-1},
\\ \label{eq:ijestimatedState} \mathsf{\hat{\boldsymbol{x}}}_{k|k}^{ij} &=\mathbf{\hat{{x}}}_{k|k-1}^{i} +\mathsf{W}_k^{ij} \boldsymbol{\nu}_{k}^{ij} ,
\\ \label{eq:ijKalstateCov} \mathsf{P}_{k|k}^{ij} & = P_{k|k-1}^{i} -\mathsf{W}_k^{ij}S_k^{ij}\mathsf{W}{_k^{ij}}\tran.
\end{align}
where \(\left(.\right)\tran \)indicates the transpose of its argument. 

Hence, defining
\begin{align}
\mu_k^{ij} & \triangleq  p {\left(M_{k}^{ij}|\mathbf{z}_{1:k}\right)},
\end{align}
we can write~\eqref{eq:GSFmergedposterior} as a GM distribution, with \({C_{v_k}}\times{C_{w_k}}\) components:
\begin{align}
\label{eq:GSFmergedposteriorSimple}
p {\left(\mathbf{x}_k | \mathbf{z}_{1:k}\right)} & = \sum \limits _{i,j} \mu_k^{ij} \mathcal{N} {\left(\mathbf{x}_k; \mathsf{\hat{\boldsymbol{x}}}_{k|k}^{ij}, \mathsf{P}_{k|k}^{ij}\right)}.
\end{align}
The coefficients \(\mu_k^{ij}\) can be evaluated as:
\begin{align}
\nonumber \mu_k^{ij}&= p {\left(M_{k}^{ij}|\mathbf{z}_k,\mathbf{z}_{1:k-1}\right)}
\\ \nonumber & = \frac{p{\left(\mathbf{z}_k |M_{k}^{ij},\mathbf{z}_{1:k-1}\right)} p{\left(M_{k}^{ij}|\mathbf{z}_{1:k-1}\right)}}{p{\left(\mathbf{z}_{k}|\mathbf{z}_{1:k-1}\right)}}
\\ \label{eq:posteriorCoeff} & = \frac{\Lambda_k^{ij}p{\left(M_{k}^{ij}|\mathbf{z}_{1:k-1}\right)}}{\sum \limits _{lm}\Lambda_k^{lm}p{\left(M_k^{lm}|\mathbf{z}_{1:k-1}\right)}},
\end{align}
where \(\Lambda_k^{ij}\) is the likelihood function and is defined as:
\begin{align}
\nonumber \Lambda_k^{ij} & \triangleq p{\left(\mathbf{z}_k |M_{k}^{ij},\mathbf{z}_{1:k-1}\right)}
\\ \label{eq:ijLikelihood} & = \mathcal{N}{\left(\mathbf{z}_k; \mathbf{\hat{z}}_{k}^{ij}, S_k^{ij}\right)},
\end{align}
and assuming that the current model is independent from the previous model we have\footnote{This assumption can be easily relaxed.}
\begin{align}
p{\left(M_{k}^{ij}|\mathbf{z}_{1:k-1}\right)} = w^i_kp^j_k.
\end{align}
Hence, we can write:
\begin{align}
\label{eq:muijComplete}
\mu_k^{ij} = \frac{w^i_kp^j_k\mathcal{N}{\left(\mathbf{z}_k; \mathbf{\hat{z}}_{k}^{ij}, S_k^{ij}\right)}}{\sum \limits _{lm} w^l_kp^m_k\mathcal{N}{\left(\mathbf{z}_k; \mathbf{\hat{z}}_{k}^{lm}, S_k^{lm}\right)}}.
\end{align}
\subsection{Reduction Schemes}
 As mentioned earlier, to avoid an exponentially growing bank size, reduction schemes should be applied to the posterior. In our work we use the less computationally complex schemes: merging all components to their moment-matching Gaussian distribution, and removing the components with smaller weights.\footnote{Other metrics can be used to determine the active model.} 
 
For the first method, the moment-matched Gaussian distribution will have the following mean and covariance matrix:
\begin{align}
\label{eq:mergedState}
\mathsf{\hat{\boldsymbol{x}}}_{k|k} = &\sum \limits _{ij} \mu_{k}^{ij}\mathsf{\hat{\boldsymbol{x}}}_{k|k}^{ij},
\\ \label{eq:mergedCov}\mathsf{P}_{k|k}  = & \sum \limits _{ij}\mu_{k}^{ij} \left( \mathsf{P}_{k|k}^{ij} \right.
\\ \label{eq:spreadOfMeans} &\left.+ \left( \mathsf{\hat{\boldsymbol{x}}}_{k|k}^{ij}- \mathsf{\hat{\boldsymbol{x}}}_{k|k}\right)\left( \mathsf{\hat{\boldsymbol{x}}}_{k|k}^{ij}- \mathsf{\hat{\boldsymbol{x}}}_{k|k}\right)\tran\right)
\\ \label{eq:mergedCovAlt}
= & \sum \limits _{ij}\mu_{k}^{ij} \left( \mathsf{P}_{k|k}^{ij} + \mathsf{\hat{\boldsymbol{x}}}_{k|k}^{ij} \mathsf{\hat{\boldsymbol{x}}}{_{k|k}^{ij}}\tran \right)- \mathsf{\hat{\boldsymbol{x}}}_{k|k}\mathsf{\hat{\boldsymbol{x}}}_{k|k}\tran.
\end{align} 

Alternatively, rather than making a soft decision about the active model, it can be determined by a hard decision. This is the approach used in~\cite{alspach_nonlinear_1972,sorenson_recursive_1971,chen_mixture_2000,sun_mixture_2004}. A simple scheme to determine the active model is to choose the component with the largest weight. Using this approach, if 
\begin{align}
ij = \arg\max_{lm}\mu_k^{lm},
\end{align}
we have
\begin{align}
\label{eq:mergeStateRemove}
\mathsf{\hat{\boldsymbol{x}}}_{k|k} = & \mathsf{\hat{\boldsymbol{x}}}_{k|k}^{ij},
\\ \label{eq:mergedCovRemove} \mathsf{P}_{k|k}  = & \mathsf{P}_{k|k}^{ij} .
\end{align}

One of the advantages of using this method over replacing the GM with its moment-matched Gaussian distribution, is that by determining the active model before evaluating the parameters of all components, computational resources can be saved. Additionally, evaluating~\eqref{eq:mergedState}--\eqref{eq:mergedCovAlt} is more computationally complex than~\eqref{eq:mergeStateRemove}--\eqref{eq:mergedCovRemove}. Besides the computational complexity, by using a soft decision approach, we could be getting drifted from the \textit{matched} estimation. Specifically, at every iteration only one model is active corresponding to the \textit{Matched} filter. By incorporating the outputs of the mismatched filters in~\eqref{eq:mergedState}--\eqref{eq:mergedCovAlt}, the estimation accuracy and precision are lost. However, if the active model, hence the Matched filter are not chosen correctly,  the soft decision method will provide better estimations. For simplicity, we refer to the first method as \textit{merge} and the second as \textit{remove}, e.g. GSF with the first method as the reduction scheme is referred to as \textit{GSF-merge} in this paper.

Additionally, throughout this paper we use the term \textit{mode-matched filter} to refer to the filter estimating the mode-conditioned state, from \(p {\left(\mathbf{x}_k |M_{k}^{ij}, \mathbf{z}_{1:k}\right)}\). The term \textit{Matched filter}, is used to denote the mode-matched filter corresponding to the known active model, \(M_k^*\). Having the information about the active model,~\eqref{eq:GSFmergedposterior} is simplified and the posterior of the Matched filter will be
\begin{align}
\label{eq:MatchedfilterPosterior}
p {\left(\mathbf{x}_k | \mathbf{z}_{1:k}\right)} & = p {\left(\mathbf{x}_k | \mathbf{z}_{1:k},M_k^* \right)}.
\end{align}
Matched filter cannot be implemented without prior knowledge about the active model and is only used for comparison purposes (see Section~\ref{sec:SimSynthetic}). 

\section{AMMSE Estimator for GM noise}
\label{sec:myCont}
In this section we evaluate the gains of individual filters, \(W_k^{ij}\) such that the trace of the covariance matrix of the bank of filters,~\eqref{eq:mergedCovAlt} is minimized. Since this covariance matrix is conditional on the measurements sequence, the evaluated gains are functions of innovations. 

\subsection{Derivation of Filter Gains}
\label{sec:DerivingGainsOptimal}
For an arbitrary filter gain \(W_k^{ij}\), the state estimation error covariance matrix for filter \(ij\) can be written as follows:\footnote{For Kalman gain this equation is simplified as in~\eqref{eq:ijKalGain}.} 
\begin{align}
\nonumber
P_{k|k}^{ij} = & P_{k|k-1}^{i} - W_k^{ij}H_kP_{k|k-1}^{i} 
\\ \label{eq:ijCovJoseph} &- P_{k|k-1}^{i}H_k\tran{W_k^{ij}}\tran+ W_k^{ij}S_k^{ij}{W_k^{ij}}\tran
\end{align}
Additionally, using~\eqref{eq:ijpredictedState} and~\eqref{eq:ijestimatedState}, we can write
\begin{align}
\nonumber
\mathbf{\hat{x}}_{k|k}^{ij}  =& F_k\mathbf{\hat{x}}_{k-1|k-1} + \mathbf{u}_k^i + W_k^{ij}\boldsymbol{\nu}_k^{ij}.
\end{align}
Using this in~\eqref{eq:mergedState}, we have:
\begin{align}
\nonumber \mathbf{\hat{x}}_{k|k} =& \sum \limits _{ij} \mu_k^{ij} \left(F_k\mathbf{\hat{x}}_{k-1|k-1} + \mathbf{u}_k^i \right) + \mathbb{S}_k
\\ \label{eq:xAMMSE} = & F_k\mathbf{\hat{x}}_{k-1|k-1} + \mathbb{U}_k + \mathbb{S}_k,
\end{align}
where
\begin{align}
\label{eq:defSS}
\mathbb{S}_k&\triangleq \sum \limits _{ij}\mu_k^{ij}W_k^{ij}\boldsymbol{\nu}_k^{ij},
\\ \mathbb{U}_k & \triangleq \sum \limits _{ij} \mu_k^{ij}\mathbf{u}_k^i.
\end{align}
Hence,
\begin{align*}
\mathbf{\hat{x}}_{k|k}^{ij} \mathbf{\hat{x}}{_{k|k}^{ij}}\tran  =& F_k\mathbf{\hat{x}}_{k-1|k-1}\mathbf{\hat{x}}_{k-1|k-1}\tran F_k^T + F_k\mathbf{\hat{x}}_{k-1|k-1}\mathbf{u}{_k^i}\tran
\\ &+F_k\mathbf{\hat{x}}_{k-1|k-1}\boldsymbol{\nu}{_k^{ij}}\tran {W_k^{ij}}\tran + \mathbf{u}_k^i \mathbf{\hat{x}}_{k-1|k-1}\tran F_k\tran
\\ & + \mathbf{u}_k^i{\mathbf{u}_k^i}\tran + \mathbf{u}_k^i\boldsymbol{\nu}{_k^{ij}}\tran{W_k^{ij}}\tran 
\\ & + W_k^{ij}\boldsymbol{\nu}{_k^{ij}}\mathbf{\hat{x}}_{k-1|k-1}\tran F_k\tran + W_k^{ij}\boldsymbol{\nu}{_k^{ij}} {\mathbf{u}_k^i}\tran 
\\&  +  W_k^{ij}\boldsymbol{\nu}{_k^{ij}} \boldsymbol{\nu}{_k^{ij}}\tran{W_k^{ij}}\tran ,
\end{align*}
and
\begin{align*}
\mathbf{\hat{x}}_{k|k}\mathbf{\hat{x}}_{k|k}\tran = & F_k\mathbf{\hat{x}}_{k-1|k-1}\mathbf{\hat{x}}_{k-1|k-1}\tran F_k\tran + F_k\mathbf{\hat{x}}_{k-1|k-1}\mathbb{U}_k\tran
\\ & + F_k\mathbf{\hat{x}}_{k-1|k-1}\mathbb{S}_k\tran + \mathbb{U}_k\mathbf{\hat{x}}_{k-1|k-1}\tran F_k\tran +\mathbb{U}_k\mathbb{U}_k\tran
\\ &  + \mathbb{U}_k\mathbb{S}_k\tran + \mathbb{S}_k\mathbf{\hat{x}}_{k-1|k-1}\tran F_k\tran + \mathbb{S}_k\mathbb{U}_k\tran +\mathbb{S}_k\mathbb{S}_k\tran.
\end{align*}
To find the optimal \(W_k^{ij}\) we need to have:
\begin{align}
\frac{\partial \mathrm{tr} \left(P_{k|k}\right)}{\partial W_k^{ij}} =& 0.
\end{align}
Thus, using the above and~\eqref{eq:ijCovJoseph} in~\eqref{eq:mergedCovAlt}, we can write:\footnote{We assume \(\mu_k^{ij}\neq 0\), since for \(\mu_k^{ij}=0\) there is no need to evaluate the parameters of filter \(ij\).}
\begin{align}
\nonumber - & P_{k|k-1}^{i}H_k\tran+W_k^{ij}S_k^{ij}
\\ \nonumber +& \mathbf{u}_k^i\boldsymbol{\nu}{_k^{ij}}\tran + W_k^{ij}\boldsymbol{\nu}_k^{ij}\boldsymbol{\nu}{_k^{ij}}\tran
\\ \label{eq:equalityConditionGains} - & \mathbb{U}_k\boldsymbol{\nu}{_k^{ij}}\tran - \mathbb{S}_k\boldsymbol{\nu}{_k^{ij}}\tran = 0
\end{align}
and we will have:
\begin{align}
\nonumber W_k^{ij} = & \left(P_{k|k-1}^{i}H_k\tran + \mathbb{U}_k\boldsymbol{\nu}{_k^{ij}}\tran - \mathbf{u}_k^i\boldsymbol{\nu}{_k^{ij}}\tran +  \mathbb{S}_k\boldsymbol{\nu}{_k^{ij}}\tran\right)
\\ \nonumber & \times \left(S_k^{ij} +\boldsymbol{\nu}_k^{ij}\boldsymbol{\nu}{_k^{ij}}\tran\right)^{-1}
\\ \label{eq:OptimalGainSimpleForm} & = \mathrm{A}_k^{ij} + \mathbb{S}_k\mathrm{B}_k^{ij},
\end{align}
where
\begin{align}
\nonumber
\mathrm{A}_k^{ij}  \triangleq & \left(P_{k|k-1}^{i}H_k\tran + \mathbb{U}_k\boldsymbol{\nu}{_k^{ij}}\tran - \mathbf{u}_k^i\boldsymbol{\nu}{_k^{ij}}\tran\right)
\\ \label{eq:Aij} & \times \left(S_k^{ij} +\boldsymbol{\nu}_k^{ij}\boldsymbol{\nu}{_k^{ij}}\tran\right)^{-1},
\\ \label{eq:Bij} \mathrm{B}_k^{ij}  \triangleq & \boldsymbol{\nu}{_k^{ij}}\tran\left(S_k^{ij} +\boldsymbol{\nu}_k^{ij}\boldsymbol{\nu}{_k^{ij}}\tran\right)^{-1}.
\end{align}
Using~\eqref{eq:defSS},~\eqref{eq:OptimalGainSimpleForm}, \(\mathbb{S}_k\) can be evaluated as follows:
\begin{align}
&\sum \limits _{ij}\mu_k^{ij}W_k^{ij}\boldsymbol{\nu}_k^{ij}  = \sum \limits _{ij}\mu_k^{ij}\mathrm{A}_k^{ij}\boldsymbol{\nu}_k^{ij} + \mathbb{S}_k\sum \limits _{ij}\mu_k^{ij}\mathrm{B}_k^{ij}\boldsymbol{\nu}_k^{ij}.
\end{align}
Since the term \(\sum \limits _{ij}\mu_k^{ij}\mathrm{B}_k^{ij}\boldsymbol{\nu}_k^{ij}\) is a scalar, we can write
\begin{align}
& \mathbb{S}_k \left( 1- \sum \limits _{ij}\mu_k^{ij}\mathrm{B}_k^{ij}\boldsymbol{\nu}_k^{ij}\right) = \sum \limits _{ij}\mu_k^{ij}\mathrm{A}_k^{ij}\boldsymbol{\nu}_k^{ij},
\end{align}
thus,
\begin{align}
\label{eq:bbsEval}
\mathbb{S}_k = & \sum \limits _{ij}\mu_k^{ij}\mathrm{A}_k^{ij}\boldsymbol{\nu}_k^{ij}\left( 1- \sum \limits _{ij}\mu_k^{ij}\mathrm{B}_k^{ij}\boldsymbol{\nu}_k^{ij}\right)^{-1}.
\end{align}
This can be used in~\eqref{eq:OptimalGainSimpleForm} to determine the optimal gain for filter \(ij\).

Having the above, the estimated state, \(\mathbf{\hat{x}}_{k|k}\) minimizing the trace of the covariance matrix of the filter can be evaluated from~\eqref{eq:xAMMSE}. However, changing the parameters of the components in the posterior, leads to larger MSE when compared with GSF, the MMSE filter.\footnote{GSF is the MMSE estimator when no reduction scheme is used and the noise distributions are GMs, rather than being \textit{approximated} by GMs. However, as shown in~\cite{ackerson_state_1970} even with mixture reduction, GSF-merge converges to the MMSE filter.} Specifically, the MSE of AMMSE filter can be evaluated as follows:
\begin{align}
\label{eq:MSE_AMMSE}
\textrm{MSE}^{\textrm{AMMSE}} \triangleq & \expectx{\left(\mathbf{x}_{k}-\mathbf{\hat{x}}_{k|k}\right)\left(\mathbf{x}_{k}-\mathbf{\hat{x}}_{k|k}\right) \tran \big| \mathbf{{z}}_{1:k}} \\
= & \mathsf{P}_{k|k}+ \left( \mathsf{\hat{\boldsymbol{x}}}_{k|k}- \mathbf{\hat{x}}_{k|k}\right)\left( \mathsf{\hat{\boldsymbol{x}}}_{k|k}-\mathbf{\hat{x}}_{k|k}\right)\tran
\end{align}
where \(\mathds{E}_\mathbf{x}{\left\lbrace g{\left(\mathbf{{x}}\right)} \right\rbrace} \) is the expected value of the function \(g{\left(\mathbf{{x}}\right)}\) with respect to the random variable \(\mathbf{{x}}\) with pdf \(p{\left(\mathbf{{x}}\right)}\), and \(\mathsf{\hat{\boldsymbol{x}}}_{k|k}\),  \(\mathsf{P}_{k|k}\) are the parameters of the MMSE filter evaluated as in~\eqref{eq:mergedState}--\eqref{eq:mergedCovAlt}.

\subsection{Comparison with GSF}
\label{sec:CompareGSFMMSE}
In this section we compare GSF and the proposed AMMSE filter in terms of computational complexity for one iteration (Section~\ref{sec:CompareComputationComplexity}) and discuss the convergence of AMMSE estimator to GSF (Section~\ref{sec:ConvergenceMMSEtoGSF}). 

\subsubsection{Computational Complexity}
\label{sec:CompareComputationComplexity}
In the first stage of Bayesian tracking, i.e. prediction, GSF and AMMSE have the same steps 
in \eqref{eq:ijpredictedState}--\eqref{eq:ijInnovationCov},\eqref{eq:ijLikelihood},\eqref{eq:muijComplete}, hence the same computational complexity. The update stage can be divided into three main operations: evaluating the gains, the component means, and covariance matrices. However, depending on the reduction scheme used, the number of times that these parameters have to be evaluated changes. Specifically, when the components are merged, the parameters of all \({C_{v_k}}{C_{w_k}}\) filters in the banks have to be evaluated, whereas when the components with smaller weights are removed, only the parameters of the filter with the maximum weight need to be computed. Moreover, merging the means and covariance matrices of the components in~\eqref{eq:mergedState}--\eqref{eq:mergedCov} is more computationally complex than simply removing the components with smaller weights in~\eqref{eq:mergeStateRemove}--\eqref{eq:mergedCovRemove}. In Table~\ref{tab:ComputationalComplexity}, we provide the computational complexity of AMMSE-merge/-remove and GSF-merge/-remove for the update stage and reduction scheme. The matrix inversions are not considered in this table for two reasons: 1) The three filters with comparable performances, i.e. AMMSE-merge/-remove and GSF-merge all have the same computational complexity in this respect, specifically, \({C_{v_k}}{C_{w_k}}\) inversions of matrices of size \(n_z \times n_z\), and although GSF-remove requires only one such inversion, its performance is not comparable to the others; 2) Even by not considering the matrix inversions, GSF-remove has the lowest computational complexity and the table serves its purpose of comparison. The detailed number of operations required for evaluating all parameters are given in Tables~\ref{tab:NumberofOprtationsforGSF}--\ref{tab:NumberofOprtationsformerge} in Appendix~\ref{sec:appCompComp}.

When merging is used as a reduction scheme, both GSF and AMMSE have the same computational complexity for evaluating the gains, component means and covariance matrices, as well as merging the moments of the individual filters. For GSF-remove, each parameter needs to be evaluated for the \textit{correct} component only. However, this is not true for AMMSE-remove, as the gains are dependent. Thus, to compute \(\mathbb{S}_k\), the parameters \(\mathrm{A}_k^{ij}\) have to be evaluated for all filters. Having \(\mathbb{S}_k\), the parameters of the component with the maximum weight can be computed with the same computational complexity as GSF-remove. Additionally, the number of matrix inversions for GSF-remove is \({C_{v_k}C_{w_k}}^{-1}\) times the number of inversions for all the other three filters. Hence, GSF-remove has the lowest computational complexity. However, as shown in Section~\ref{sec:SimulationResults} this is at the cost of losing estimation precision and accuracy. The other three filters, i.e. AMMSE-merge/-remove and GSF-merge show similar performance with AMMSE-remove requiring the least number of operations. 
\begin{table*}[!t]
\centering
\caption{Computational Complexity of GSF-merge/-remove, and AMMSE-merge/-remove}
\label{tab:ComputationalComplexity}
\begin{tabular}{l|c|c}
 Filter & Parameter & Computational complexity \\
\hline
\multirow{4}{*}{GSF-merge} & \(\left\lbrace\mathsf{W}_k^{ij}; 1 \leq i \leq {C_{v_k}}, 1\leq j \leq {C_{w_k}} \right\rbrace \) &  \(O \left({C_{v_k}}{C_{w_k}}n_x^2n_z\right)\) \\
& \(\left\lbrace\mathsf{P}_{k|k}^{ij}; 1 \leq i \leq {C_{v_k}}, 1\leq j \leq {C_{w_k}} \right\rbrace\)  & \(O \left({C_{v_k}}{C_{w_k}}n_x^2n_z\right)\) \\
& \(\left\lbrace\mathsf{\hat{\boldsymbol{x}}}_{k|k}^{ij}; 1 \leq i \leq {C_{v_k}}, 1\leq j \leq {C_{w_k}} \right\rbrace\) & \(O \left({C_{v_k}}{C_{w_k}}n_x n_z\right)\) \\
& \(\mathsf{\hat{\boldsymbol{x}}}_{k|k},\mathsf{P}_{k|k}\) & \(O\left({C_{v_k}}{C_{w_k}n_x^2}\right)\) \\
\hline
\multirow{4}{*}{AMMSE-merge} & \(\left\lbrace{W}_k^{ij}; 1 \leq i \leq {C_{v_k}}, 1\leq j \leq {C_{w_k}} \right\rbrace \) &  \(O \left({C_{v_k}}{C_{w_k}}n_x^2n_z\right)\) \\
& \(\left\lbrace{P}_{k|k}^{ij}; 1 \leq i \leq {C_{v_k}}, 1\leq j \leq {C_{w_k}} \right\rbrace\)  & \(O \left({C_{v_k}}{C_{w_k}}n_x^2n_z\right)\) \\
& \(\left\lbrace\mathbf{\hat{x}}_{k|k}^{ij}; 1 \leq i \leq {C_{v_k}}, 1\leq j \leq {C_{w_k}} \right\rbrace\) & \(O \left({C_{v_k}}{C_{w_k}}n_x n_z\right)\) \\
& \(\mathbf{\hat{x}}_{k|k}^{ij},{P}_{k|k}\) & \(O\left({C_{v_k}}{C_{w_k}n_x^2}\right)\) \\
\hline
\multirow{3}{*}{GSF-remove} & \(\left\lbrace\mathsf{W}_k^{ij}; ij = \arg\max_{lm}\mu_k^{lm}\right\rbrace\) &  \(O \left(n_x^2n_z\right)\) \\
& \(\left\lbrace\mathsf{P}_{k|k}^{ij}; ij = \arg\max_{lm}\mu_k^{lm}\right\rbrace\)  & \(O \left(n_x^2n_z\right)\) \\
& \(\left\lbrace\mathsf{\hat{\boldsymbol{x}}}_{k|k}^{ij}; ij = \arg\max_{lm}\mu_k^{lm}\right\rbrace\) & \(O \left(n_x n_z\right)\) \\
\hline
\multirow{3}{*}{AMMSE-remove} & \(\mathbb{S}_k\) & \(O \left({C_{v_k}}{C_{w_k}}n_x^2n_z\right)\)  \\
& \(\lbrace P_{k|k}^{ij}; ij = \arg\max_{lm}\mu_k^{lm}\rbrace\)  & \(O \left(n_x^2n_z\right)\)\\
& \(\lbrace \mathbf{\hat{x}}_{k|k}^{ij}; ij = \arg\max_{lm}\mu_k^{lm}\rbrace\) & \(O \left(n_xn_z\right)\)\\
\end{tabular}
\end{table*}

\subsubsection{Convergence to MMSE Estimator}
\label{sec:ConvergenceMMSEtoGSF}
At iteration \(k\), the AMMSE state estimation, \(\mathbf{\hat{x}}_{k|k}\), converges in distribution to the MMSE state estimation provided by GSF, \(\mathsf{\hat{\boldsymbol{x}}}_{k|k}\). Using~\eqref{eq:MSE_AMMSE}, the convergence of the state estimations yields the convergence of the MSEs. Hence, the MSE of the AMMSE filter converges to the MSE of GSF-merge. Since it is not easily feasible to provide an analytical proof for all cases
due to the matrix inversions in the evaluation of gains, we prove the convergence for two limit cases in Appendix~\ref{sec:appConvergence}, and use Kolmogrov-Smirnov test for the other cases in Section~\ref{sec:SimulationResults}.

In the analytical proof in Appendix~\ref{sec:appConvergence}, the convergence of state estimations is proved by showing the convergence of the parameters of the GM posterior, namely the coefficients, \(\mu_k^{ij}\), the means, \(\mathbf{\hat{x}}_{k|k}^{ij}\), and the covariance matrices, \(P_{k|k}^{ij}\), of the components. Since these parameters are dependent on the innovations of individual filters, \(\boldsymbol{\nu}_{k}^{ij}\), two limit cases are considered: when the distance between innovations approaches zero and when it goes to infinity. The first case applies to a posterior with highly overlapping components, where the likelihoods of all models are close to one. Contrarily, when the distance between innovations approaches infinity, the likelihood of the active model is close to one and all the other components have negligible likelihoods. In other words, the active model can be well determined by using the coefficients, \(\mu_k^{ij}\), of the GM posterior.

For general GM noise models, the analytical proof is not straight forward, due to the matrix inversions in the gains, \(\left(S_k^{ij} +\boldsymbol{\nu}_k^{ij}\boldsymbol{\nu}{_k^{ij}}\tran\right)^{-1}\) and \({S_k^{ij}}^{-1}\). Hence, in Section~\ref{sec:SimulationResults}, we provide Kolmogrov-Smirnov (KS) statistic for the distributions, \(p {\left(\mathbf{\hat{x}}_{k|k}\right)}\) and \(p {\left(\mathsf{\hat{\boldsymbol{x}}}_{k|k}\right)}\) under different types of GM noise parameters. Specifically, we use GM noise models with different separations between components, and by running simulations on synthetically generated data with these noise models, we generate samples from the distributions \(p {\left(\mathbf{\hat{x}}_{k|k}\right)}\) and \(p {\left(\mathsf{\hat{\boldsymbol{x}}}_{k|k}\right)}\). Using the KS test on these data, the hypothesis that the two data samples belong to the same distribution, is accepted at \(\% 95\) confidence interval.

\section{Simulation Results}
\label{sec:SimulationResults}
We consider two simulation scenarios: In the first scenario, we use synthetically generated process and measurement noise processes, whereas for the second simulation scenario we gather experimental data from  an indoor localization system with ultra-wideband (UWB) sensors. 
To be consistent, we use the same process and measurement equations for both scenarios and they only differ in the noise distributions.

For both scenarios, we consider an indoor localization problem, in a 2D setting, and track the position in each direction independently assuming noise distributions with time-invariant statistics. The state vector contains the position and the velocity, but only noisy information about the position is observable and measured. Hence, in each direction we have \(n_x =2\), \(n_z=1\), and
\begin{align}
F_k = & \begin{bmatrix}
1 & \Delta t_{k} \\
0 & 1
\end{bmatrix},\\
H_k = & \begin{bmatrix}
1 & 0
\end{bmatrix},
\end{align}
where \(\Delta t_{k}\) is the time interval between the measurements \(z_{k-1}\) and \(z_k\). In our localization system, the time intervals between measurements are multiples of \(0.1080\)~s. Hence, in our synthetic setting, we use \(\Delta t_k=0.1080\)~s for all iterations, \(k\).

We use a random walk velocity motion model,
\begin{align}
\mathbf{v}_{k} = v_k \times \begin{bmatrix}
\Delta t_{k} \\
1
\end{bmatrix},
\end{align}
where \(v_k\) is a univariate GM random variable with \(C_{v_k}\) components, with \(\left\lbrace {u}^i_k,1 \leq i \leq C_{v_k}\right\rbrace\) the component means, \(\left\lbrace {\sigma^i}^2_k,1 \leq i \leq C_{v_k}\right\rbrace\) the component variances. Hence, in~\eqref{eq:priorDist} we have:
\begin{align}
\forall i; \; 1 \leq i \leq C_{v_k}, \; & \mathbf{u}^i_k = {u}^i_k \times \begin{bmatrix}
\Delta t_{k} \\
1
\end{bmatrix},\\
& Q^i_k = {\sigma^i_k}^2 \times \begin{bmatrix}
\Delta t_{k}^2 &  \Delta t_{k}\\
\Delta t_{k} & 1
\end{bmatrix}.
\end{align}

For the experimental setup, the noise distributions, \(p{\left(v_k\right)}\) and \(p{\left(w_k\right)}\) are estimated using the data gathered from the UWB sensors. For the synthetically generated data, we assume the same distribution for process and measurement noise, i.e. \(p{\left(v_k\right)}=p{\left(w_k\right)}\).

The following filtering schemes are used and compared:
\begin{enumerate}
\item Kalman filter (KF)
\item GSF-merge/-remove
\item AMMSE-merge/-remove
\item Matched filter (for synthetically generated data)
\end{enumerate}
The performance metrics that we used for analyzing the estimation accuracy and precision are 
root-mean-square error (RMSE) and circular error probable (CEP), respectively. Denoting the location estimation error at iteration \(k\) by \(\epsilon_k\), the RMSE and CEP are defined as:
\begin{align}
\textrm{RMSE} \triangleq & \left({\frac{1}{N}\sum \limits_{k=1}^{N}{\epsilon_k}^2}\right)^{1/2}, \\
\textrm{CEP} \triangleq & F^{-1}_{|\epsilon|} \left(0.5\right),
\end{align}
where \(N\) is the total number of iterations and \(F^{-1}_{|\epsilon|}\) represents the inverse cumulative distribution function of error evaluated over the whole experiment. 
\subsection{Synthetic data}
\label{sec:SimSynthetic}
For the synthetically generated data, the process and measurement noises are assumed to be i.i.d. samples from one of the following GM models:
\begin{itemize}
{\setlength\itemindent{35pt}\item[\textbf{Model 1:}] Symmetric distribution with the components all having the same coefficients.}
{\setlength\itemindent{35pt}\item[\textbf{Model 2:}] Symmetric distribution with components possibly having different weights.}
{\setlength\itemindent{35pt}\item[\textbf{Model 3:}] Asymmetric distribution.}
\end{itemize}
The coefficients of the components are chosen such that the noises are zero mean. In our simulations, we assumed GM distributions with 5 components, each having a variance of \(1\), i.e. we have
\begin{align}
 C_{v_k}  = C_{w_k} & = 5, \\
\forall i; \;  1 \leq i \leq 5, & \; {\sigma^i_k}^2 = 1 ,
\\ \forall j; \;  1 \leq j \leq 5, & \;  R^j_k = 1.
\end{align}

The coefficients, \(\mathbf{w}\), and the means, \(\mathbf{m}\), of the components are given in Table~\ref{tab:GMModelsparams}, where
\begin{align}
\mathbf{w}  & = \begin{bmatrix}
w_k^1, \cdots, w_k^5
\end{bmatrix} = \begin{bmatrix}
p_k^1,\cdots , p_k^5
\end{bmatrix}, \\
\mathbf{m} & = \begin{bmatrix}
b_k^1,\cdots , b_k^5
\end{bmatrix} = \begin{bmatrix}
u_k^1,\cdots , u_k^5
\end{bmatrix}. 
\end{align} 
The parameter \(c\) is multiplied by the means to change the distance between components.

\begin{table}[!t]
\centering
\caption{The parameters of the GM models used for generating synthetic data}
\label{tab:GMModelsparams}
\begin{tabular}{c l}
\hline
\textbf{Model 1} & \(\mathbf{w} = \left[0.2,0.2,0.2,0.2,0.2\right]\) \\
& \(\mathbf{m}=c \left[-50,-30,0,30,50\right]\) \\
\hline
\textbf{Model 2} & \(\mathbf{w} = \left[0.1,0.1,0.6,0.1,0.1\right]\) \\
& \(\mathbf{m}=c\left[-50,-30,0,30,50\right]\) \\
\hline
\textbf{Model 3} & \(\mathbf{w} = \left[0.5,0.1,0.1,0.1,0.2\right]\) \\
& \(\mathbf{m}=c\left[-50,10,30,50,80\right]\) \\
\hline
\end{tabular}
\end{table}

In addition to the above models, to provide more general simulation results we also used the noise models from ~\cite[Section~V-B--C]{bilik_mmse-based_2010} corresponding to equal means and unequal covariance matrices. The first scenario corresponds to a maneuvering target tracking problem with GM process noise and Gaussian measurement noise. The second example corresponds to a system with Gaussian process noise and glint measurement noise. The parameters of the used models are provided in Table~\ref{tab:GMModelsparamsBilik}. In this table, we use \(\prescript{}{}{[a^i]}^{n}_{1}\) to denote \([a^1,\cdots,a^n]\).
\begin{table}[!t]
\centering
\caption{The parameters of the two scenarios from~\cite{bilik_mmse-based_2010}}
\label{tab:GMModelsparamsBilik}
\begin{tabular}{c l}
\hline
\textbf{Maneuvering target} & \(C_{v_k} = 2\) \\
& \(\prescript{}{}{[w_k^i]}^{2}_{1} = [0.8,0.2]\) \\
&  \(\prescript{}{}{[u_k^i]}^{2}_{1} = [0,	0]\) \\
& \(\prescript{}{}{[{\sigma^i}^2_k]}^{2}_{1} = [0.01,1]\)\\
& \(C_{w_k}= 1\), \(b_k = 0\), \(R_k = 0.1\) \\
\hline
\textbf{Glint measurement noise} & \(C_{v_k}= 1\), \(u_k = 0\), \({\sigma}^2_k = 1\) \\
& \(C_{w_k} = 2\) \\
& \(\prescript{}{}{[p_k^j]}^{2}_{1} = [0.1,0.9]\) \\
&  \(\prescript{}{}{[b_k^j]}^{2}_{1} = [0,	0]\) \\
& \(\prescript{}{}{R_k^j}^{2}_{1} = [0.01,1]\)\\
\end{tabular}
\end{table}

\begin{figure}[!t]
\centering
\subfloat[RMSE]{\includegraphics[width=3.0in]{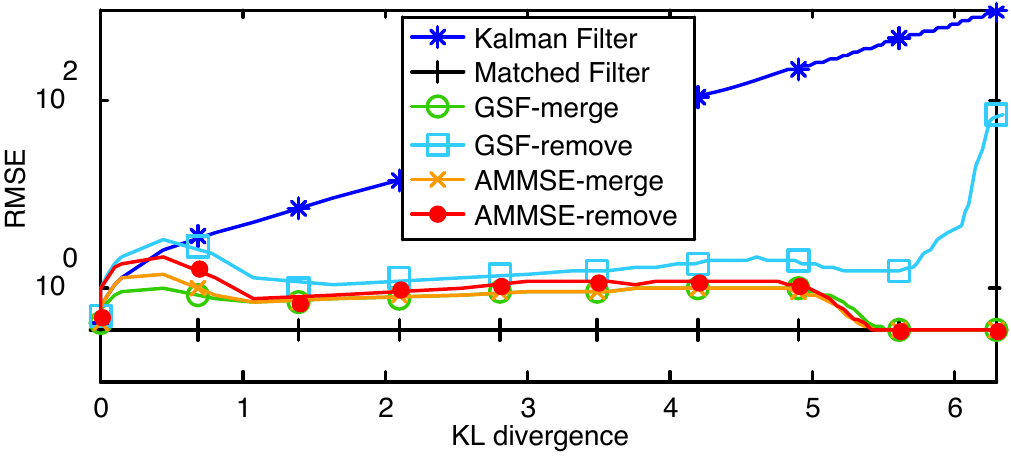}}%
\hfill
\subfloat[CEP]{\includegraphics[width=3.0in]{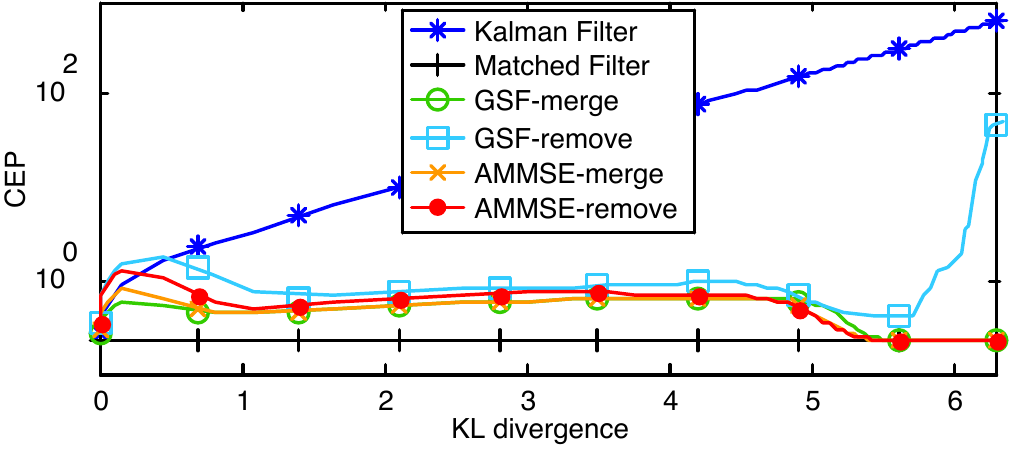}}
\caption{RMSE and CEP for the synthetic data generated using Model 1 vs. KL divergence between the noise distribution and the moment-matched Gaussian pdf.}
\label{fig:RMSECEPM1}
\end{figure}

\begin{figure}[!t]
\centering
\subfloat[RMSE]{\includegraphics[width=3.0in]{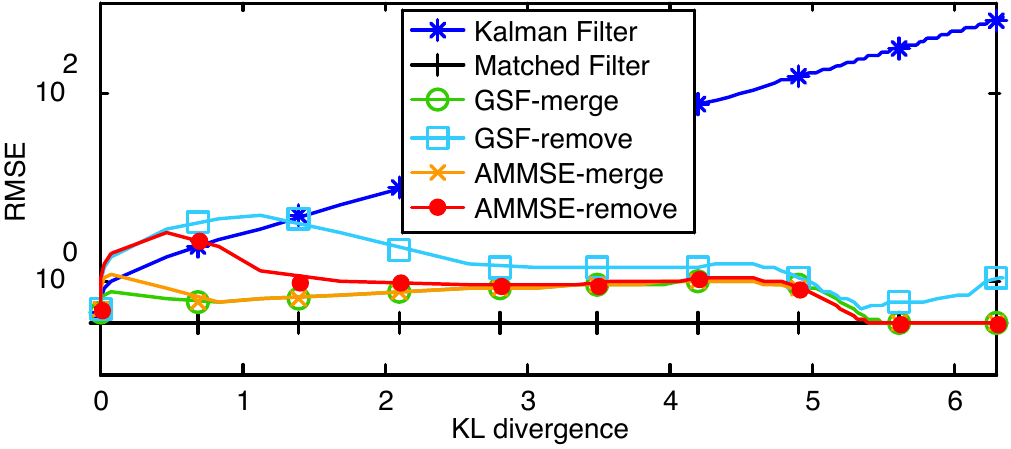}}%
\hfill
\subfloat[CEP]{\includegraphics[width=3.0in]{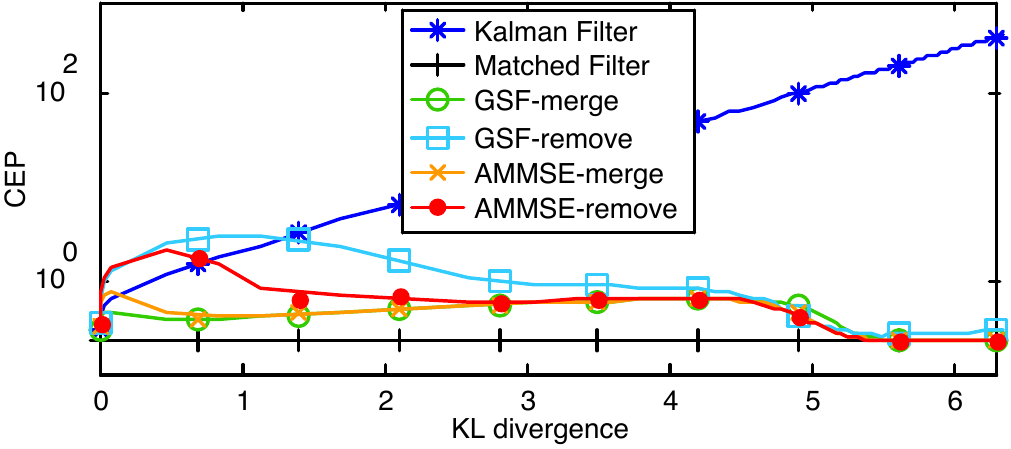}}
\caption{RMSE and CEP for the synthetic data generated using Model 2 vs. KL divergence between the noise distribution and the moment-matched Gaussian pdf.}
\label{fig:RMSECEPM2}
\end{figure}

\begin{figure}[!t]
\centering
\subfloat[RMSE]{\includegraphics[width=3.0in]{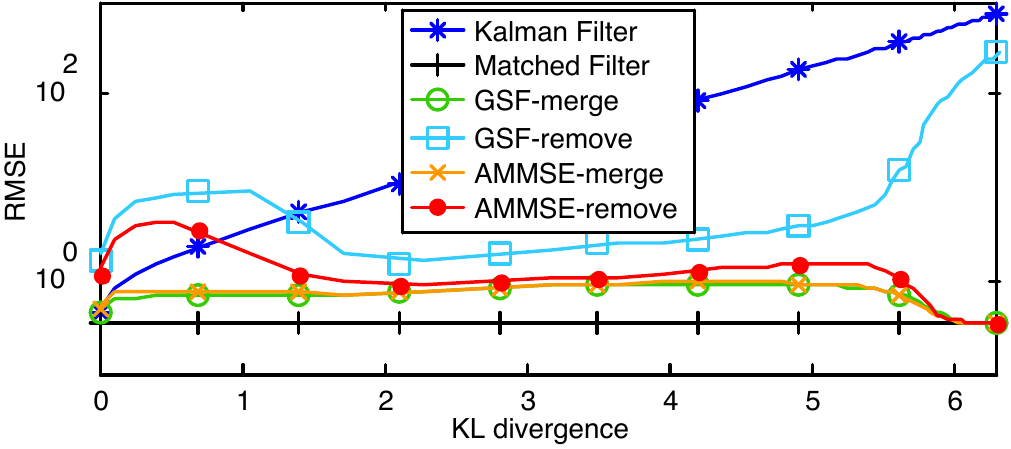}}%
\hfill
\subfloat[CEP]{\includegraphics[width=3.0in]{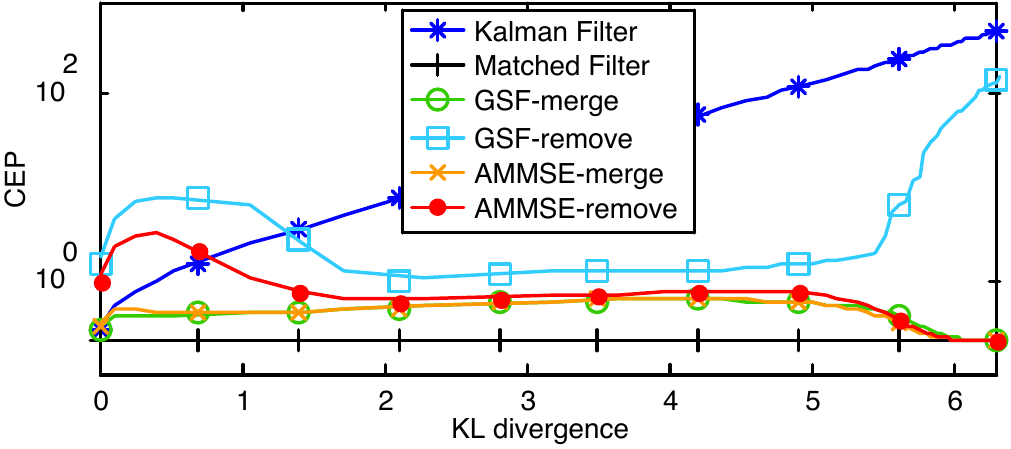}}
\caption{RMSE and CEP for the synthetic data generated using Model 3 vs. KL divergence between the noise distribution and the moment-matched Gaussian pdf.}
\label{fig:RMSECEPM3}
\end{figure}

Using the GM noise models in Table~\ref{tab:GMModelsparams}--\ref{tab:GMModelsparamsBilik}, we generated the measurements \(z_k\) and estimated the state with Kalman filter, GSF-merge/-remove, and AMMSE-merge/-remove, as well as the Matched filter. To do this, we label the generated data by the active noise models in effect, and use this information to choose the correct filter in the bank to achieve the Matched filter. To further investigate the effect of multi-modality on the performance of these filters, we vary the parameter \(c\) for the noise models in Table~\ref{tab:GMModelsparams} and evaluate the state estimate, RMSE, and CEP for all filters. For each value of \(c\), \(1000\) Monte-Carlo runs are used to estimate the RMSE, and CEP at \( \%95\) confidence interval. We also approximate the KL divergence between the GM noise distribution and their corresponding moment-matched Gaussian density for each value of \(c\). The state estimations evaluated in these simulations are then used as samples from the pdfs \(p \left(\mathbf{\hat{x}}_{k|k}\right)\), and \(p \left(\mathsf{\hat{\boldsymbol{x}}}_{k|k}\right)\) to approximate the Kolmogrov-Smirnov (KS) statistic and test the hypothesis that the two sets of samples have the same distribution. The results for GM noise distribution with Model 1, 2, and 3 in Table~\ref{tab:GMModelsparams} are provided in Fig.~\ref{fig:RMSECEPM1}, Fig.~\ref{fig:RMSECEPM2}, and Fig.~\ref{fig:RMSECEPM3}, respectively. The results for the maneuvering target tracking scenario and the system with glint measurement noise with the parameters defined in Table~\ref{tab:GMModelsparamsBilik} are provided in Table~\ref{tab:BilikResults}. These results show the RMSE and CEP of the filters for different values of KL divergence between the used noise model and its moment-matched Gaussian distribution. As can be expected, as the divergence between the noise distribution and its corresponding fitted Gaussian increases, the performance of Kalman filter drops both in terms of RMSE and CEP. This is due to the fact that Kalman filter is best suited for systems with Gaussian distributions, and it fails to provide good estimations for multi-modal noise models. Additionally, since the component variances of noise distributions are the same and by varying \(c\) only the separation between the components changes, the performance of the Matched filter remains the same for all values of divergence. However, for GSF-merge/-remove, and AMMSE-merge/-remove, the performance changes when the KL divergence increases.

With all noise models, the performance of all filters are close to the Matched filter for small KL divergences (as shown in Section~\ref{sec:ConvergenceMMSEtoGSF}). However,  as the divergence increases, both RMSE and CEP increase for all filters, until they reach their maximum. This is due to the fact that with an increase in the separation between the noise components, hence, the KL divergence with the corresponding moment-matched Gaussian distribution, the overlap between the components is decreasing. However, this decrease is not enough for the filters to correctly find the active model by the weights of components. In other words, the posterior is multi-modal, hence it cannot be well approximated by a single Gaussian. However, the overlap between the components of the GM posterior leads to drifting from the Matched filter estimation. Further increase in the divergence results in improved performance for AMMSE-merge/-remove and GSF-merge until they converge to the Matched filter (as shown in Section~\ref{sec:ConvergenceMMSEtoGSF}). However, this is not the case for GSF-remove. Specifically, as the KL divergence increases, the RMSE and CEP of GSF-remove decrease until they reach their minimum. But further increase in the KL divergence results in increased RMSE and CEP for this filter. This is because with GSF-remove the correct choice of the active component is of particular importance. Although, with the increase in KL divergence the component with the maximum weight represents the active model most of the times, when it fails and a wrong component is chosen, there will be a larger error due to the increased distance between the component corresponding to the active model and the other components. 

For all models, the performances of GSF-merge and AMMSE-merge are very close, especially, for KL divergence values greater than 1. To further investigate the relationship between these two filters, we used the state estimations, \(\mathbf{\hat{x}}_{k|k}\), and \(\mathsf{\hat{\boldsymbol{x}}}_{k|k}\) to test the hypothesis that they come from the same distribution. Using KS test, this hypothesis is accepted at \(\%95\) confidence interval for all noise models and KL divergences. Additionally, since GSF-merge is not the MMSE filter due to the reduction of GM posteriors, we also carried out the two-sample KS hypothesis tests on the state estimations from AMMSE-merge and the true state. The hypotheses that the samples have the same distribution are accepted at \(\%95\) confidence interval for all noise models for KL divergences greater than 1.Moreover, to test the variances of the two filters we used Ansari-Bradley\footnote{Since the posteriors are not Gaussian the chi-square tests for normalized estimation error squares (NEES) cannot be applied.} test with the null hypothesis that the variances of these two sample sets are equal. The null hypotheses for Ansari-Bradley tests were also accepted for all noise models and KL divergences greater than \(0.04\) at \(\%95\) confidence interval.

The RMSE and CEP of AMMSE-remove are always smaller or equal to the RMSE and CEP of GSF-remove. This is due to the fact that with AMMSE, the means of components are evaluated such that the trace of total state estimation error covariance matrix including the spread of means in~\eqref{eq:spreadOfMeans} is minimized. Contrarily, in GSF the component means, \(\mathsf{\hat{\boldsymbol{x}}}_{k|k}^{ij}\) are minimizing the trace of the covariance matrix of each individual filter \(\mathsf{P}_{k|k}^{ij}\). Hence, the component means are closer in AMMSE filter. Consequently, when removing the components with smaller weights as a reduction scheme, the performance (both accuracy and precision) is better. 

Based on the results, it is evident that when the noise distributions are far from Gaussian (the KL divergence between the noise distribution and the corresponding moment-matched Gaussian density is large), GSF-merge, and AMMSE-merge/-remove perform similarly regardless of the shape of the noise model in terms of symmetry. This is particularly important, since among these filters which achieve comparable estimation accuracy and precision, AMMSE-remove requires the least number of operations by avoiding the evaluation of \(P_{k|k}^{ij}\) (as shown in Section~\ref{sec:CompareComputationComplexity}).

%

%

\begin{table}[!h]
\centering
\caption{Simulation results for maneuvering target tracking and glint measurement noise}
\label{tab:BilikResults}
\begin{tabular}{l|c|c|c|c|}
\cline{2-5}
               & \multicolumn{2}{c|}{\begin{tabular}[c]{@{}c@{}}Maneuvering target\\ KL div = 0.0155\end{tabular}} & \multicolumn{2}{c|}{\begin{tabular}[c]{@{}c@{}}Glint noise\\ KL div = 0.0042\end{tabular}} \\ \cline{2-5} 
               & RMSE                                                 & CEP                                                 & RMSE                                               & CEP                                               \\ \hline
Kalman         & 0.1204                                               & 0.0802                                              & 0.3356                                             & 0.2191                                            \\ \hline
Matched filter & 0.1184                                               & 0.0780                                              & 0.2274                                             & 0.1208                                            \\ \hline
GSF-merge      & 0.1204                                               & 0.0802                                              & 0.3224                                             & 0.2007                                            \\ \hline
GSF-remove     & 0.1866                                               & 0.1325                                              & 0.3354                                             & 0.2195                                            \\ \hline
AMMSE-merge    & 0.1204                                               & 0.0802                                              & 0.3312                                             & 0.2158                                            \\ \hline
AMMSE-remove   & 0.1860                                               & 0.1306                                              & 0.3343                                             & 0.2187                                            \\ \hline
\end{tabular}
\end{table}

%


%

\subsection{Experimental Data: Indoor Positioning System}
\label{sec:simExperimental}
In this section we provide the simulation results of the experimental data gathered from an indoor positioning system with off-the-shelf \textit{Ubisense} UWB location sensors. In our setting, we have four UWB receivers, four stationary objects and a mobile target. A UWB \textit{tag} is attached to each stationary and mobile object and sends UWB pulses to the four receivers. The system estimates the location of these tags by computing the time and angle difference of arrival of their UWB pulses. 
%

\begin{figure}[!t]
\centering
\includegraphics[width=3.0in]{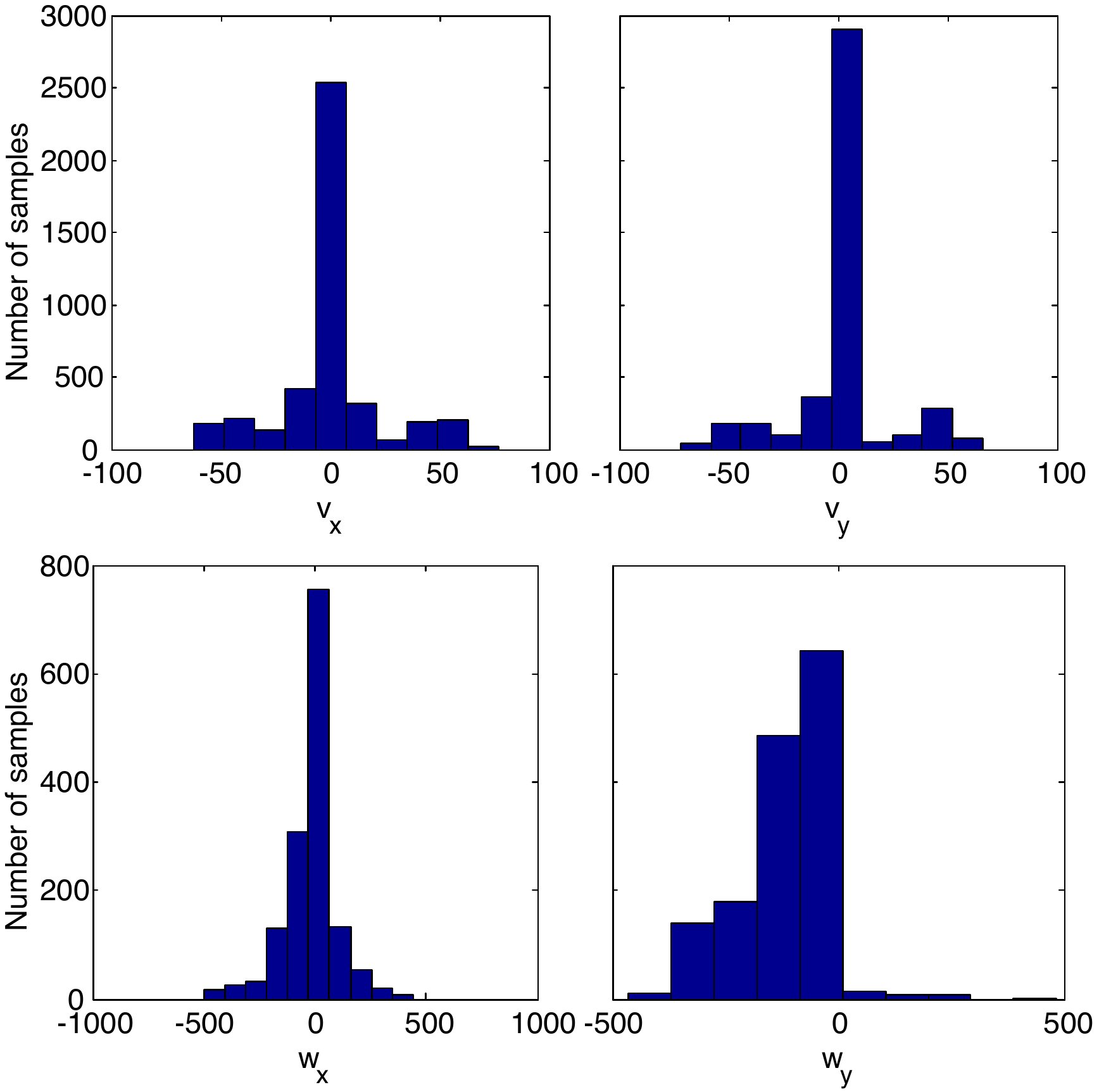}
\caption{Histogram of process and measurement noise in \(x\) and \(y\) directions}
\label{fig:AllNoisepdf}
\end{figure}

The location information of the stationary objects is used for approximating the measurement noise distribution by GMs. The histograms of measurement noise in \(x\) and \(y\) directions are given in Fig.~\ref{fig:AllNoisepdf}.
%
%
To find the process noise pdfs, we used the ground truth about the location of the mobile target. The experiment is carried out \(20\) times with the mobile target moving on a predefined trajectory.
The histograms of the process noise are depicted in Fig.~\ref{fig:AllNoisepdf} for \(x\) and \(y\) directions.

%

With GM approximations for process and measurement noise, the position is estimated by post-processing the sensor measurements. The approximated parameters of the GM noise distributions in both directions, and also the KL divergence with their corresponding moment-matched Gaussian distributions are given in Table~\ref{tab:KLdiv}. In this table, we use \(\prescript{}{}{[a^i]}^{n}_{1}\) to denote \([a^1,\cdots,a^n]\).

\begin{table}[!t]
\centering
\caption{The parameters of the GM noise distributions}
\label{tab:KLdiv}
\begin{tabular}{@{\hspace{0.1cm}}l@{\hspace{0.1cm}}|c@{\hspace{0.1cm}}|@{\hspace{0.1cm}}c@{\hspace{0.1cm}}|l}
                        &  \multirow{2}{*}{Noise}    & KL  & \multirow{2}{*}{Parameters} \\
                        & & divergence & \\
                       \hline
\normalsize 
{\multirow{12}{*}{\(x\)}} & \multirow{6}{*}{\(v_k\)} & \multirow{6}{*}{\(0.4253\)} & {\(C_{v_k}=3\)}
\\[1mm]
& & & {\(\prescript{}{}{[w_k^i]}^{3}_{1} = [0.13,0.77,0.099]\)}       \\[1mm]
& & &  {\(\prescript{}{}{[u_k^i]}^{3}_{1} = [-41.44,	0.51,	49.79]\)} \\[1mm]
& & &  {\(\prescript{}{}{[{\sigma^i}^2_k]}^{3}_{1} = [148.24,48.38,	83.75]\)} \\[1mm]
\cline{2-4}
& \multirow{6}{*}{\(w_k\)} & \multirow{6}{*}{\(0.1759\)} & {\(C_{w_k}=3\)} \\[1mm]
& & &  {\(\prescript{}{}{[p_k^j]}^{3}_{1} = [0.07,	0.85,	0.08]\)}       \\[1mm]
& & & {\(\prescript{}{}{[b_k^j]}^{3}_{1} = [-300.01,	-17.06,	207.37]\)} \\[1mm]
& & & {\(\prescript{}{}{[R_k^j]}^{3}_{1} = [8163.20,	3611.99,	5677.21]\)} \\[1mm]
\hline                       
\multirow{16}{*}{\(y\)} & \multirow{8}{*}{\(v_k\)} & \multirow{8}{*}{\(1.1971\)}       & \(C_{v_k}=9\) \\[1mm]
& & &  \(\prescript{}{}{[w_k^i]}^{9}_{1} = [ 0.01,	0.06,	0.03, 0.03,\)       \\
& & & \( \quad \quad \quad \quad \; 	0.72,	0.04 ,  0.02,	0.06,	0.03 ]\)\\
& & &  \(\prescript{}{}{[u_k^i]}^{9}_{1} = [-63.38,	-48.73,	-35.65,	-17.40,\) \\ 
& & &\(\quad \quad \quad \quad \; 	-0.32,	9.52,	30.09,	44.24,	54.35]\) \\
& & &  \(\prescript{}{}{[{\sigma^i}^2_k]}^{9}_{1} = [24.34,	21.53,	18.18,	23.62,\) \\ 
& & &\(\quad \quad \quad \quad \; 	3.13,	12.16,	18.81,	12.96,	15.44]\) \\
\cline{2-4}
& \multirow{6}{*}{\(w_k\)} & \multirow{6}{*}{\(0.0200\)} & \(C_{w_k}=2\) \\[1mm]
& & &  \(\prescript{}{}{[p_k^j]}^{2}_{1} = [0.98,	0.02]\)\\[1mm]
& & & \(\prescript{}{}{[b_k^j]}^{2}_{1} = [-125.93,	147.25]\) \\[1mm]
& & & \(\prescript{}{}{[R_k^j]}^{2}_{1} = [8500.19,10809.10]\) \\[1mm]
\end{tabular}
\end{table}

\begin{figure}[!t]
\centering
\subfloat[\(x\) direction]{\includegraphics[width=3.0in]{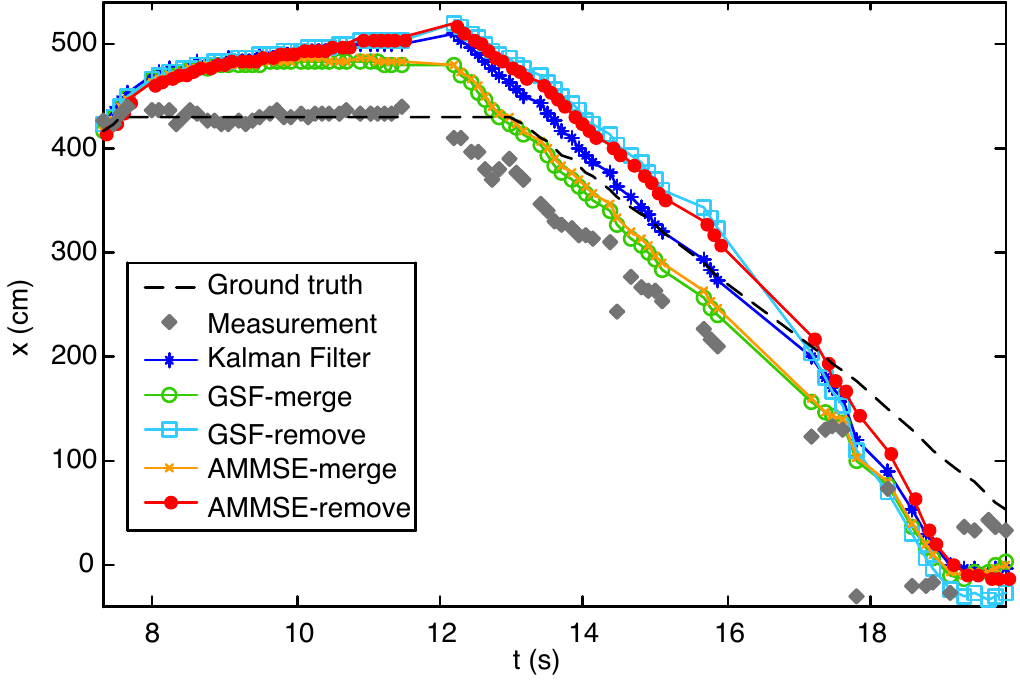}}%
\hfill
\subfloat[\(y\) direction]{\includegraphics[width=3.0in]{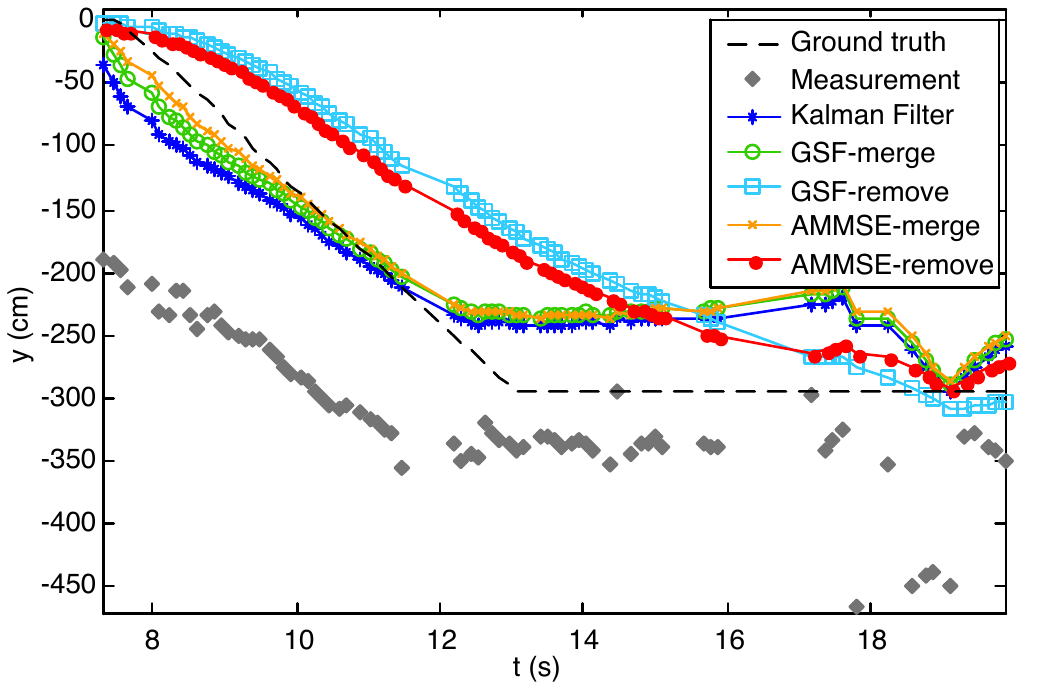}}
\caption{Location estimation in \(x\) and \(y\) directions}
\label{fig:estimatedState}
\end{figure}

Fig.~\ref{fig:estimatedState} shows the location estimations in \(x\) and \(y\) directions for one of our experiments. To better compare the performance of different filtering schemes, in Table~\ref{tab:experimentalResults} we provide the RMSE and CEP in each direction. These values are the average of RMSE and CEP over all our experiments. Since the noise distributions are very close to their moment-matched Gaussian distribution (Table~\ref{tab:KLdiv}), the RMSE and CEP of Kalman estimations are very close to the RMSE and CEP of GSF-merge and AMMSE-merge. However, by minimizing the trace of the total state estimation error covariance matrix and decreasing the spread of means, AMMSE provides the best results in both directions and the performance of AMMSE-remove is better than GSF-remove.

%

\begin{table}[!t]
\centering
\caption{simulation results for experimental data}
\label{tab:experimentalResults}
\begin{tabular}{lr|r|r|r|}
& \multicolumn{2}{c|}{x} & \multicolumn{2}{c}{y} \\ 
\cline{2-5} 
 \multicolumn{1}{l|}{} & \multicolumn{1}{c|}{RMSE}       & \multicolumn{1}{c|}{CEP}       & \multicolumn{1}{c|}{RMSE}       & \multicolumn{1}{c|}{CEP}        \\ \hline
\multicolumn{1}{|l|}{Kalman}       & 88.1850	&	66.3306	&	73.0512	&	60.4984    \\ \hline
\multicolumn{1}{|l|}{GSF-merge}    & 85.7485	&	55.2261	&	74.3982	&	61.7358    \\ \hline
\multicolumn{1}{|l|}{GSF-remove}    & 110.4006	&	91.7458	&	75.1093	&	48.6651    \\ \hline
\multicolumn{1}{|l|}{AMMSE-merge}  & \textbf{80.7388}	&	\textbf{52.2901}	&	\textbf{71.3662}	&	56.7287     \\ \hline
\multicolumn{1}{|l|}{AMMSE-remove} & 101.8026	&	83.2669	&	72.0491	&	\textbf{44.7549}
   \\ \hline
\end{tabular}
\end{table}

\begin{figure}[!t]
\centering
\subfloat[GSF]{\includegraphics[width=3.0in]{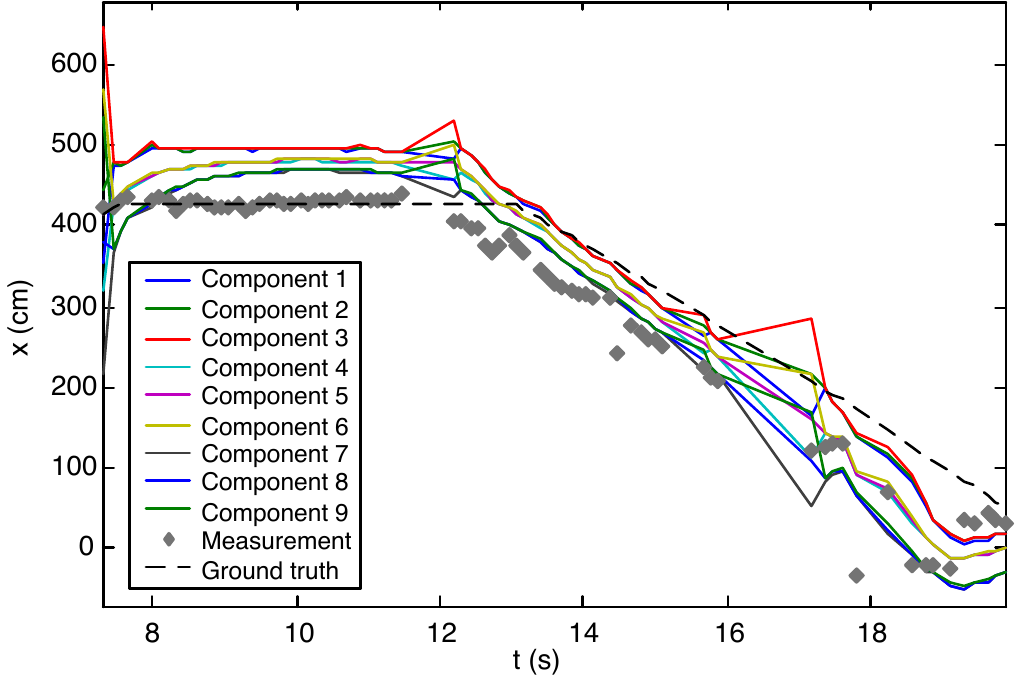}}
\hfill
\subfloat[AMMSE]{\includegraphics[width=3.0in]{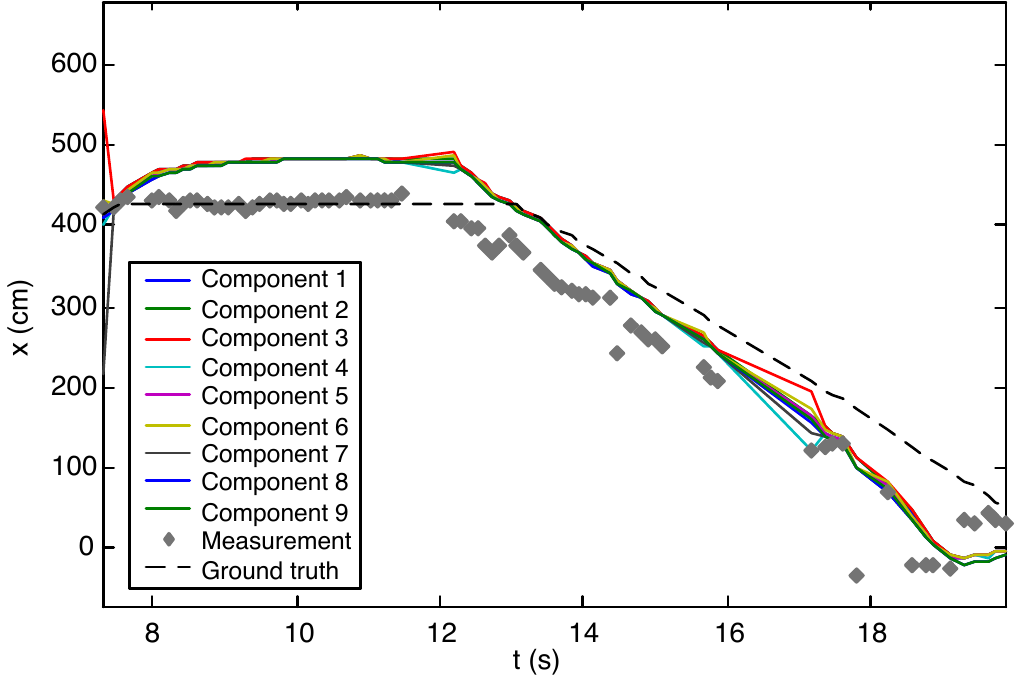}}
\caption{Component centers for GSF and AMMSE in \(x\) direction}
\label{fig:ClusterCenters}
\end{figure}

To better depict the difference in spread of means of GSF and AMMSE, their component means are depicted in Fig.~\ref{fig:ClusterCenters}. This figure shows the component means for one of our experiments in \(x\) direction. As shown in this figure, the spread of means in GSF is larger than that of AMMSE. This is especially evident at times \(t = 12.19\)~s and \(t = 17.17\)~s.

%
%
%


%

\section{Conclusion}
\label{sec:Conclusion}
In this paper, we propose an approximate MMSE (AMMSE) state estimator, for linear dynamic systems with Gaussian Mixture (GM) noise. For this purpose, we used a bank of Kalman filters with adjusted gains to track the models corresponding to the different components of the GM noises. This is done by minimizing the trace of the total state estimation error covariance matrix, including the individual filters covariance matrices and the spread of their means. Hence, comparing with Gaussian Sum Filter (GSF) which minimizes the trace of the individual filters covariance matrices by using parallel Kalman filters, our proposed method has a smaller spread of means, and is more robust to removing components. Specifically, we have shown through simulations that unlike GSF, the performance of the proposed AMMSE filter does not change when instead of merging all components, we reduce the number of components by taking the component with the maximum weight. This is specifically important for applications which require lower computational complexities. We have also shown that the distributions of state estimations with GSF and AMMSE filter converge at two limit cases: when the distance between the GM components approach zero and infinity. For the other cases, this is tested with Kolmogrov-Smirnov test on the state estimations of the two filters.


%

\appendices
\section{Number of Required Operations for the Filters}
\label{sec:appCompComp}
In this section we provide the number of required operations for GSF (Table~\ref{tab:NumberofOprtationsforGSF}), Approximate MMSE (Table~\ref{tab:NumberofOprtationsforMMSE}) and merging the means and covariance matrices of individual filters (Table~\ref{tab:NumberofOprtationsformerge}). In the first column of these tables, we indicate the required operations for evaluating each parameter. The last column represents the number of times that a certain parameter has to be evaluated. For instance, evaluating the gains has to be done for every filter whereas the value \(\mathbb{S}_k\) is only evaluated once for every iteration. However, with remove as the reduction scheme, the number of filters becomes one, and each row in Tables~\ref{tab:NumberofOprtationsforGSF}--\ref{tab:NumberofOprtationsforMMSE} is evaluated only once. 

To further compare the two reduction algorithms in terms of computational complexity, we provide the number of required operations for merging the components in Table~\ref{tab:NumberofOprtationsformerge}. For simplicity, we  use the GSF parameters in this table, however, it applies to both filters.

In these tables, we assume that multiplying \(A_{a \times b}\) by  \(B_{b \times c}\), where the subscripts represent the sizes of the matrices, requires \(ac(b-1)\) additions and \(abc\) multiplications. Additionally, we assume that the transposition of matrices requires no operations and the inversion of a scalar is similar to a multiplication.

\begin{table*}[!t]
\centering
\caption{Number of Required Operations for Update Stage of GSF}
\label{tab:NumberofOprtationsforGSF}
\begin{tabular}{ll|c|c|c}
 \multicolumn{2}{r|}{Operation} & additions & multiplications & Frequency \\
\hline
\multirow{2}{*}{\(\mathsf{W}_k^{ij}:\)} & \(P_{k|k-1}^{ij}H_k\tran \) & \(n_xn_z\left(n_x-1\right)\) & \(n_x^2n_z\) & \multirow{2}{*} {filter} \\
& \(\mathsf{W}_k^{ij}\)  & \(n_xn_z\left(n_z-1\right)\) & \(n_xn_z^2\) & \\
\hline
\multirow{3}{*}{\(\mathsf{P}_{k|k}^{ij}:\)} & \(W_k^{ij}S_k^{ij}\) & \(n_xn_z\left(n_z-1\right)\) & \(n_xn_z^2\) & \multirow{3}{*} {filter}\\
& \(W_k^{ij}S_k^{ij}{W_k^{ij}}\tran \) & \(n_x^2\left(n_z-1\right)\) & \(n_x^2n_z\)& \\
& \(P_{k|k-1}^{ij} \) & \(n_x^2\)& \(0\) & \\
\hline
\multirow{2}{*}{\(\mathsf{\hat{\boldsymbol{x}}}_{k|k}^{ij}:\)} & \(W_k^{ij}\boldsymbol{\nu}_k^{ij}\) & \(n_x \left(n_z-1\right)\) & \(n_xn_z\) & \multirow{2}{*}{filter}\\
& \(\mathsf{\hat{\boldsymbol{x}}}_{k|k}^{ij}\) & \(n_x\) & \(0\) & \\
\end{tabular}
\end{table*}

\begin{table*}[!t]
\centering
\caption{Number of Required Operations for Update Stage of Approximate MMSE filter}
\label{tab:NumberofOprtationsforMMSE} 
\begin{tabular}{ll|c|c|c}
 \multicolumn{2}{r|}{Operation} & additions & multiplications & Frequency \\
\hline
 \multicolumn{2}{l|}{\(\mathbb{U}_k\)} & \({C_{v_k}}{C_{w_k}}-1\) & \(n_x{C_{v_k}}{C_{w_k}}\) & iteration \\
 \hline
\multirow{6}{*} {\(\mathrm{A}_k^{ij}:\)} & \(P_{k|k-1}^{ij}H_k\tran \) & \(n_xn_z\left(n_x-1\right)\) & \(n_x^2n_z\) & \multirow{6}{*} {filter} \\ 
& \(\mathbb{U}_k-\mathbf{u}_k^i\) & \(n_x\) & \(0\) & \\
& \(\left(\mathbb{U}_k-\mathbf{u}_k^i\right)\boldsymbol{\nu}{_k^{ij}}\tran \) & \(0\) & \(n_xn_z\) & \\
& \(P_{k|k-1}^{ij}H_k\tran  +\left(\mathbb{U}_k-\mathbf{u}_k^i\right)\boldsymbol{\nu}{_k^{ij}}\tran \) & \(n_xn_z\) & \(0\) & \\
& \(S_k^{ij} +\boldsymbol{\nu}_k^{ij}\boldsymbol{\nu}{_k^{ij}}\tran \) & \(n_z^2\) & \(n_z^2\) & \\
& \(\mathrm{A}_k^{ij}\) &  \(n_xn_z\left(n_z-1\right)\) &  \(n_xn_z^2\) & \\
\hline
\multicolumn{2}{l|}{\(\mathrm{B}_k^{ij}\)} & \(n_z\left(n_z-1\right)\) & \(n_z^2\) & filter \\
\hline
\multirow{6}{*}{\(\mathbb{S}_k:\)} & \(\mathrm{A}_k^{ij}\boldsymbol{\nu}_k^{ij}\)& \({C_{v_k}}{C_{w_k}}n_x\left(n_z-1\right)\) & \({C_{v_k}}{C_{w_k}}n_xn_z\) & \multirow{6}{*} {iteration} \\
& \(\sum \limits _{ij}\mu_k^{ij}\mathrm{A}_k^{ij}\boldsymbol{\nu}_k^{ij}\) & \(\left({C_{v_k}}{C_{w_k}}-1\right)n_x\) & \({C_{v_k}}{C_{w_k}}n_x\) & \\
& \(\mathrm{B}_k^{ij}\boldsymbol{\nu}_k^{ij}\) & \({C_{v_k}}{C_{w_k}}\left(n_z-1\right)\) & \({C_{v_k}}{C_{w_k}}n_z\) &  \\
& \(\sum \limits _{ij}\mu_k^{ij}\mathrm{B}_k^{ij}\boldsymbol{\nu}_k^{ij}\) & \({C_{v_k}}{C_{w_k}}-1\) & \({C_{v_k}}{C_{w_k}}\) &  \\
& \(\left( 1- \sum \limits _{ij}\mu_k^{ij}\mathrm{B}_k^{ij}\boldsymbol{\nu}_k^{ij}\right)^{-1}\) & \(1\) &\(1\) &\\
& \(\mathbb{S}_k\) & \(0\) & \(n_x\) & \\
\hline
\multirow{2}{*}{\(W_k^{ij}:\)} & \(\mathbb{S}_k\mathrm{B}_k^{ij}\)& \(0\) & \(n_xn_z\) & \multirow{2}{*} {filter} \\
& \(\mathrm{A}_k^{ij} + \mathbb{S}_k\mathrm{B}_k^{ij}\)  & \(n_xn_z\) & \(0\) & \\
\hline
\multirow{5}{*}{\(P_{k|k}^{ij}:\)} &\(H_kP_{k|k-1}^{ij}\)&\(n_xn_z\left(n_x-1\right)\) & \(n_x^2n_z\) & \multirow{5}{*} {filter} \\
& \(W_k^{ij}H_kP_{k|k-1}^{ij}\) & \(n_x^2\left(n_z-1\right)\) & \(n_x^2n_z\) & \\
& \(W_k^{ij}S_k^{ij}\) & \(n_xn_z\left(n_z-1\right)\) & \(n_xn_z^2\) & \\
& \(W_k^{ij}S_k^{ij}{W_k^{ij}}\tran \) & \(n_x^2\left(n_z-1\right)\) & \(n_x^2n_z\)& \\
& \(P_{k|k}^{ij} \) & \(3n_x^2\)& \(0\) & \\
\hline
\multirow{2}{*}{\(\mathbf{\hat{x}}_{k|k}^{ij}:\)} & \(W_k^{ij}\boldsymbol{\nu}_k^{ij}\) & \(n_x \left(n_z-1\right)\) & \(n_xn_z\) & \multirow{2}{*}{filter}\\
& \(\mathbf{\hat{x}}_{k|k}^{ij}\) & \(n_x\) & \(0\) & \\
\end{tabular}
\end{table*}

\begin{table*}[!t]
\centering
\caption{Number of Required Operations for merging}
\label{tab:NumberofOprtationsformerge}
\begin{tabular}{ll|c|c|c}
 \multicolumn{2}{r|}{Operation} & additions & multiplications & Frequency \\
\hline
 \multicolumn{2}{l|}{\(\mathsf{\hat{\boldsymbol{x}}}_{k|k}\)} & \({C_{v_k}}{C_{w_k}}-1\) & \(n_x{C_{v_k}}{C_{w_k}}\) & iteration \\
\hline
\multirow{4}{*}{\(\mathsf{P}_{k|k}:\)} & \(\left(\mathsf{\hat{\boldsymbol{x}}}_{k|k}-\mathsf{\hat{\boldsymbol{x}}}_{k|k}^{ij}\right)\) & \(n_x\) & \(0\) & \multirow{4}{*} {iteration}\\
& \(\left(\mathsf{\hat{\boldsymbol{x}}}_{k|k}-\mathsf{\hat{\boldsymbol{x}}}_{k|k}^{ij}\right)\left(\mathsf{\hat{\boldsymbol{x}}}_{k|k}-\mathsf{\hat{\boldsymbol{x}}}_{k|k}^{ij}\right)\tran \) & \(0\) & \(n_x^2\)& \\
& \(\mathsf{P}_{k|k}^{ij}+\left(\mathsf{\hat{\boldsymbol{x}}}_{k|k}-\mathsf{\hat{\boldsymbol{x}}}_{k|k}^{ij}\right)\left(\mathsf{\hat{\boldsymbol{x}}}_{k|k}-\mathsf{\hat{\boldsymbol{x}}}_{k|k}^{ij}\right)\tran \) & \(n_x^2\)& \(0\) & \\
& \(\mathsf{P}_{k|k}\) & \({C_{v_k}}{C_{w_k}}-1\) & \(n_x^2{C_{v_k}}{C_{w_k}}\) & \\
\end{tabular}
\end{table*}

\section{Proof of Convergence}
\label{sec:appConvergence}
In this section we show that the estimated GM posteriors of AMMSE and GSF converge for two limit cases for the distance between innovations of individual filters. The convergence is proved by showing the convergence of the parameters of the GM posteriors.

\begin{enumerate}[(a)]
\item When the distances between innovations approach zero, i.e.
\begin{align}
\nonumber \forall  i\neq l, j\neq m; & 1 \leq i,l \leq {C_{v_k}}, 1 \leq j,m \leq {C_{w_k}},\\ \label{eq:InnovationsConverge} &\left\|\boldsymbol{\nu}{_k^{{ij}}}-\boldsymbol{\nu}{_k^{{lm}}}\right\| \rightarrow 0.
\end{align}
In this case, all the innovations are converging to the same value,
\begin{align}
\forall  i,j; & 1 \leq i \leq {C_{v_k}}, 1 \leq j \leq {C_{w_k}}, \quad \boldsymbol{\nu}{_k^{{ij}}} \rightarrow \boldsymbol{\nu}{_k^{*}}.
\end{align}
Hence, we have
\begin{align}
\mathbb{S}_k& \rightarrow \sum \limits _{ij}\mu_k^{ij}W_k^{ij}\boldsymbol{\nu}_k^{*}.
\end{align}

Now, using~\eqref{eq:equalityConditionGains}, we can write:
\begin{align}
\label{eq:equalityConditionGainsConverge} - & \sum \limits _{ij}\mu_k^{ij} P_{k|k-1}^{i}H_k\tran + \sum \limits _{ij}\mu_k^{ij} W_k^{ij}S_k^{ij} \rightarrow 0,
\end{align}
which holds for GSF gains, \(\mathsf{W}_k^{ij}\) in~\eqref{eq:ijKalGain}. But since there is only one unique solution for the set of equalities given in~\eqref{eq:equalityConditionGains} for all \(i,j\), we have
\begin{align}
W_k^{ij} \rightarrow \mathsf{W}_k^{ij},
\end{align}
and consequently,
\begin{align}
\hat{\mathbf{x}}_{k|k}^{ij} \rightarrow &  \mathsf{\hat{\boldsymbol{x}}}_{k|k}^{ij},
\\ {P}_{k|k}^{ij} \rightarrow & \mathsf{P}_{k|k}^{ij}.
\end{align}
\item When the distances between innovations approach infinity\footnote{We have assumed that the filters (GSF/AMMSE) are stable and have bounded error. Specifically, by choosing the wrong components, the state estimation error can increase for all individual filters in the bank leading to all the innovations approaching infinity. This is not the case considered here.}, i.e.
\begin{align}
\nonumber \forall  i\neq l, j\neq m; & 1 \leq i,l \leq {C_{v_k}}, 1 \leq j,m \leq {C_{w_k}},\\ \label{eq:InnovationsDiverge} &\left\|\boldsymbol{\nu}{_k^{{ij}}}-\boldsymbol{\nu}{_k^{{lm}}}\right\| \rightarrow \infty.
\end{align}
\end{enumerate}
In this case, GSF and AMMSE filter converge to the Matched filter. Specifically, if we denote the active model by \(M_k^* \), we will have
\begin{align}
\boldsymbol{\nu}{_k^{*}}\rightarrow 0,
\end{align}
hence,
\begin{align}
\Lambda_k^{*} \rightarrow 1.
\end{align}
On the other hand, for all the other models \(\left\lbrace M_{k}^{lm}; 1 \leq l \leq {C_{v_k}}, 1\leq m \leq {C_{w_k}}, M_{k}^{lm}\neq M_{k}^{*} \right\rbrace\), the innovations \(\boldsymbol{\nu}{_k^{lm}}\) increase and approach infinity, hence we will have
\begin{align}
\Lambda_k^{lm} \rightarrow 0.
\end{align}
Thus, using~\eqref{eq:posteriorCoeff}, we will have
\begin{align}
\mu_k^{*} & \rightarrow 1, \\
\mu_k^{lm} & \rightarrow 0 .
\end{align}
Hence, after merging\footnote{If reduction is done by removing the components with smaller weights, as in~\eqref{eq:mergeStateRemove}--\eqref{eq:mergedCovRemove}, the two sides are equal.}
\begin{align}
\mathsf{\hat{\boldsymbol{x}}}_{k|k} \rightarrow &  \mathsf{\hat{\boldsymbol{x}}}_{k|k}^{*}, \hat{\mathbf{x}}_{k|k} \rightarrow \hat{\mathbf{x}}_{k|k}^{*},
\\ \mathsf{P}_{k|k}  \rightarrow &  \mathsf{P}_{k|k}^{*}, {P}_{k|k}  \rightarrow  {P}_{k|k}^{*}.
\end{align}
Moreover, using~\eqref{eq:Aij}--\eqref{eq:Bij}, we will have
\begin{align}
\mathrm{A}_k^{*} & \rightarrow \mathsf{W}_k^{ij}, \\
\mathrm{B}_k^{*} & \rightarrow 0,
\end{align}
whereas for the non-matching models, \(\left\lbrace M_{k}^{lm}; 1 \leq l \leq {C_{v_k}}, 1\leq m \leq {C_{w_k}}, M_{k}^{lm}\neq M_{k}^{*} \right\rbrace\), 
\begin{align}
\mathrm{A}_k^{lm} & \rightarrow 0, \\
\mathrm{B}_k^{lm} & \rightarrow 0.
\end{align}
Thus,
\begin{align}
\hat{\mathbf{x}}_{k|k}^{*} \rightarrow &  \mathsf{\hat{\boldsymbol{x}}}_{k|k}^{*},
\\ {P}_{k|k}^{*} \rightarrow & \mathsf{P}_{k|k}^{*}.
\end{align}


%


\section*{Acknowledgment}

This work was partly supported by the Natural Sciences and Engineering Research Council (NSERC) and industrial and government partners, through the Healthcare Support through Information Technology Enhancements (hSITE) Strategic Research Network.

\ifCLASSOPTIONcaptionsoff
  \newpage
\fi



%

\bibliographystyle{IEEEtran}
\bibliography{KFBank}

%

%






\end{document}